\RequirePackage{lineno}
\documentclass[aps,prx,10pt,twocolumn,superscriptaddress,longbibliography]{revtex4-2}

\usepackage{graphicx}
\usepackage{amssymb,amsmath,amsfonts,gensymb}
\usepackage{bm,hyperref}
\usepackage{color}
\usepackage{ulem}
\usepackage{bbm}
\usepackage{subfigure}
\usepackage{dcolumn}
\usepackage{mathtools}
\usepackage{pifont}

\newcommand{\ket}[1]{\vert #1 \rangle}

\newcommand{\bk}{\mathbf{k}}
\newcommand{\bq}{\mathbf{q}}
\newcommand{\btk}{\widetilde{\mathbf{k}}}
\newcommand{\btq}{\widetilde{\mathbf{q}}}
\newcommand{\br}{\mathbf{r}}
\newcommand{\bG}{\mathbf{G}}
\newcommand{\bQ}{\mathbf{Q}}
\newcommand{\cop}{\hat{c}}

\def\I{\uppercase\expandafter{\romannumeral 1}}
\def\II{\uppercase\expandafter{\romannumeral 2}}
\def\III{{\uppercase\expandafter{\romannumeral 3}}}
\def\IV{{\uppercase\expandafter{\romannumeral 4}}}
\def\V{{\uppercase\expandafter{\romannumeral 5}}}
\def\VI{{\uppercase\expandafter{\romannumeral 6}}}
\def\VII{{\uppercase\expandafter{\romannumeral 7}}}

\def\k{\mathbf{k}}

\def\br{\mathbf{r}}

\begin{document}

\title{Correlation stabilized anomalous Hall crystal in bilayer graphene}

\author{Zhongqing Guo}
\affiliation{School of Physical Science and Technology, ShanghaiTech University, Shanghai 201210, China}

\author{Jianpeng Liu}
\email{liujp@shanghaitech.edu.cn}
\affiliation{School of Physical Science and Technology, ShanghaiTech University, Shanghai 201210, China}
\affiliation{ShanghaiTech Laboratory for Topological Physics, ShanghaiTech University, Shanghai 201210, China}
\affiliation{Liaoning Academy of Materials, Shenyang 110167, China}

\bibliographystyle{apsrev4-2}

\begin{abstract} 
When the charge density is sufficiently low, interacting two-dimensional electron gas (2DEG) would undergo a phase transition from homogeneous Fermi liquid to an electronic crystal state, known as Wigner crystal. Besides conventional 2DEG, various topological fermionic excitations may also be realized in 2D materials. For example, ``high-order" Dirac fermions exhibiting nontrivial Berry phases may approximately characterize the low-energy excitations in  rhombohedral multilayer graphene (RMG). In this work, we develop a beyond-mean-field theoretical framework to study the interacting ground states and single-particle excitations in slightly charge-doped RMG under vertical electric field. We find that transitions from Fermi liquid to  trivial Wigner-crystal states would occur at critical carrier density $\sim 10^{10}\,\textrm{cm}^{-2}$ for all $n$-layer RMG (with $n=2, 3, 4, 5, 6$) which are approximately described by $n$-order Dirac-fermion models. 
Most saliently, using a more realistic modeling of bilayer graphene including trigonal warping effects, we find that an anomalous Hall crystal state with spontaneous quantized anomalous Hall conductivity would emerge when the carrier density is below $\sim 1\times 10^{11}\,\text{cm}^{-2}$, and  it becomes the unique ground state over trivial Wigner crystal when the  density is further lower. Counter intuitively, such topological anomalous Hall crystal becomes more stable than the trivial Wigner crystal due to the lower correlation energy gained from dynamical charge fluctuations, which is beyond mean-field description. 
 Our work suggests that slightly carrier-doped bilayer graphene is one of the most promising candidates to realize anomalous Hall crystal. Moreover, the method developed in this work can be readily applied to other interacting 2D systems including moir\'e superlattices.
\end{abstract} 

\maketitle

\section{Introduction}
\begin{figure*}[htb]
	\centering
	\includegraphics[width=6in]{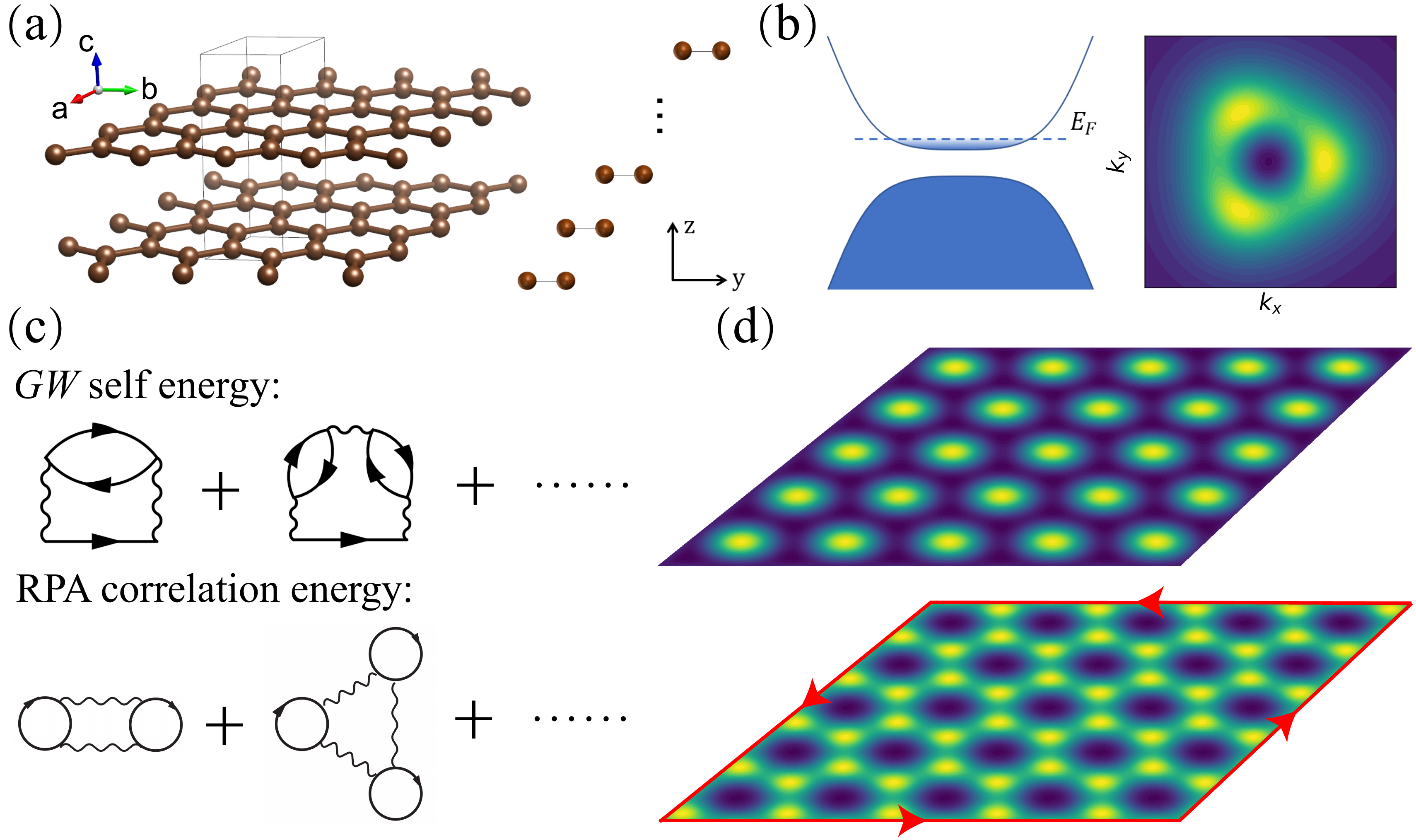}
	\caption{(a) Schematic illustration of lattice structure of bilayer graphene and RMG. (b) Left panel: schematic illustration of band structure of slightly charge-doped RMG under finite vertical electric field. Right panel: Berry curvature distribution of conduction band in bilayer graphene, including trigonal warping effects. (c) Upper panel: schematic illustration of Feynman diagrams for $GW$ self-energy corrections. Lower panel: Feynman diagrams for RPA correlation energy. The solid lines with arrows indicate single-particle Green's function, and wiggly lines denote $e$-$e$ Coulomb interactions. (d) Schematic illustration of trivial Wigner crystal (upper panel) and topological AHC (lower panel). }
\label{fig:0}
\end{figure*}

When the carrier density is sufficiently low,  interacting 2DEG system would undergo a transition from  gapless Fermi liquid (FL) state to charge-gapped Wigner crystal (WC) state \cite{wigner-pr34,eva-wigner-prl88,wigner-exp-prl99,wc-hf-prb03,needs-wigner-qmc-prl09,philips-nanoletter-18,wang-wigner-nature20,Regan2020,li-wigner-nature21,mose2-wc-nature21,signature-wc-nature21,yazdani-wc-nature24}. The latter spontaneously break spatial translational symmetry, forming a quantum electronic lattice with the lattice constant uniquely defined by the carrier density.  Besides conventional 2DEG, various two-dimensional Dirac-fermion models may also be realized in condensed matter systems. It is well known that massless Dirac fermions with linear energy-band dispersion can emerge in monolayer graphene \cite{graphene-science-04,graphene-rmp}. Moreover, even ``$n$-order" Dirac fermions with  energy-momentum dispersion $E_{\k}\sim k^n$ may approximately describe the low-energy single-particle excitations in $n$-layer graphene with rhombohedral stacking \cite{min-multilayer-08}, the  lattice structure of which is schematically shown in Fig.~\ref{fig:0}(a). When these gapless ``$n$-order" Dirac fermions are gapped out, e.g., by vertical displacement field, they would acquire non-vanishing Berry curvatures which further endow nontrivial topological properties to these Dirac fermions, as schematically shown in the right panel of Fig.~\ref{fig:0}(b). 
It is then intriguing to ask what would be the interacting ground states and characteristic single-particle excitations when the density of these topological Dirac fermions is low, as can be realized in slightly charge-doped RMG.

Several fundamental questions need to be answered for such charge-doped RMG. First, is there any WC transition happening when the carrier density in the RMG is sufficiently low?  
Second, would there be any topological properties associated with the presumable WC state due to non-vanishing Berry curvatures of the system? Previous mean-field calculations suggest that a type of topological Wigner-crystal state with nonzero Chern number, dubbed as ``anomalous Hall crystal" (AHC) state can be realized in  RMG  when the number of layers is greater than 3 \cite{dong-AHC-prl24,zhang-ahc-prl24,tan-AHC-prx24,dong-AHC-prb24}. However, it is well known that mean-field treatment drastically overestimates the tendency of Wigner crystallization \cite{needs-wigner-qmc-prl09,rapisarda-dqmc-1996,wc-hf-prb03}. It is thus crucial to study the fate of  AHC state under quantum fluctuation effects.
 Lastly, what would be the characteristics of the single-particle spectra of the RMG under slight charge doping, both in FL state and the (presumable) Wigner-crystal state? 
 
To answer these questions,  we develop a ``$GW$+RPA" framework to study both the  ground-state properties and the single-particle excitation spectra of generic interacting 2D systems. To be specific, we first calculate the ground-state energy and single-particle spectra of the system under Hartree-Fock approximation, where two trial wavefunctions have been used: one is the WC state that spontaneously breaks translational symmetry, and the other is the FL state. Then, we calculate the frequency dependent $GW$ single-particle self energy \cite{hedin-gw-pr65,aryasetiawan-gw-rpp98,reining-gw-wires18,golze-gw-fc19} (schematically shown in the left panel of Fig.~\ref{fig:0}(c)), which gives more accurate descriptions to the single-particle excitation spectra for  both WC  and FL states. Since Hartree-Fock calculations neglect correlation energy,  we further calculate correlation energy with random phase approximation (RPA) \cite{bohm-rpa-pr53,gellmann-rpa-pr57} based on the $GW$ single-particle spectra	 (see right panel of Fig.~\ref{fig:0}(c)).
We apply such ``$GW$+RPA" method to charge-doped, $n$-layer  RMG (with $n=2, 3, 4, 5, 6$), first in the simplified framework of $n$-order Dirac-fermion models.
We find that WC transitions with topologically trivial WC ground states would occur  at critical carrier density $\sim 1\times 10^{10}\,\textrm{cm}^{-2}$ for  all of these systems.  For $n=2$ Dirac-fermion model which approximately describing the low-energy physics of Bernal bilayer graphene (BLG),  an anomalous Hall crystal (AHC) state with Chern-number 1 becomes the ground state when the carrier density is below $\sim 2.3\times 10^{10}\,\text{cm}^{-2}$ thanks to the more concentrated Berry-curvature distribution in momentum space. While such AHC state becomes unstable under dynamical charge-fluctuation effects when $n\geq 3$ due to the more spreaded momentum-space distribution of Berry curvatures.   Most saliently, using a more realistic modeling of  bilayer graphene taking into account trigonal warping effects, we find that an AHC state with Chern-number 1 has lower energy than FL state when the electron density is below $\sim 1.3\times 10^{11}\,\text{cm}^{-2}$. Moreover,  the AHC becomes the unique ground state over both FL and trivial WC when the  density is lower than $\sim 2\times 10 ^{10}\,\textrm{cm}^{-2}$. 
Counter intuitively, such topological AHC is stabilized over trivial WC due to correlation effects rather than exchange effects: it is favored over the trivial WC due to lower correlation energy gained from dynamical charge fluctuations, which is entirely beyond mean-field description. The AHC is characterized by a real-space charge density distribution forming an emergent honeycomb lattice, which  exhibits ground-state currents winding around the regions of charge-density peaks. 
Furthermore, plasmaron-like excitation features are obtained in the single-particle excitation spectra of the FL states for  all the charge-doped RMG systems.

\section{$GW$+RPA methods}
To study the WC transition in both conventional 2DEG and RMG systems, we adopt the $GW$+RPA framework, which accounts for electron correlation effects beyond mean-field approximation. Compared with mean-field approach such as Hartree-Fock (HF) method, or ``Hartree-Fock+RPA" method (to be discussed below), the ``$GW$+RPA" approach enables a more accurate calculation of single-particle excitation spectra and ground-state energy, offering deeper insights into the potential transitions between FL and WC states.

\subsection{Coulomb interactions and Hartree-Fock approximation}
We consider the long-range Coulomb interaction Hamiltonian which applies to both conventional 2DEG and $n$-order Dirac fermion systems, expressed as:
\begin{equation}
	H_C = \frac{1}{2N_s} \sum_{\lambda, \lambda'} \sum_{\bk, \bk', \bq} \,V(\bq)\, \hat{c}^{\dagger}_{\bk+\bq,\lambda} \hat{c}^{\dagger}_{\bk'-\bq,\lambda'} \hat{c}_{\bk',\lambda'} \hat{c}_{\bk,\lambda},
	\label{eq:coulomb}
\end{equation}
where $N_s$ is the total number of unit cells (of presumable WC state) in the system, $\bk$ and $\bq$ represent the wave vectors expanded around some low-energy valleys, and $\lambda \equiv (\mu, \alpha, \sigma)$ is a composite index denoting the (possible) valley,  sublattice, and spin degrees of freedom. The  Coulomb interaction is in the double-gate screened form, given by $V(\bq) = e^2 \tanh(|\bq| d_s)/(2 \Omega_0 \epsilon_r \epsilon_0 |\bq|)$, where $d_s=40\,$nm represents the screening length, $\Omega_0$ is the area of a unit cell, $\epsilon_r$ is the relative dielectric constant, and $\epsilon_0$ is the vacuum permittivity.
Initially, we solve the interaction Hamiltonian using the self-consistent unrestricted HF approximation in a plane-wave basis set \cite{supp_info}. Then, we compare the total energies $E_{\text{total}}=E_{\text{kinetic}}+E_{\text{HF}}$ of the WC states and FL states, and determine the ground state at the Hartree-Fock level. Here $E_{\text{kinetic}}$ is the kinetic energy, and $E_{\text{HF}}$ is the HF energy including both Hartree and exchange energies.  

\subsection{$GW+$RPA method} 
\label{methods:gwrpa}
HF treamtment neglects electron correlation effects,  thus tends to favor symmetry-breaking states such as  WC. To improve upon this, we introduce the correlation energy through RPA, which captures collective electron-hole excitations and counterbalances the HF bias toward symmetry-breaking state. The RPA correlation energy is expressed as \cite{fetter-book,bohm-rpa-pr53,gellmann-rpa-pr57,ren-rpa-2012}:
\begin{equation}
	E_c^{\text{RPA}} = \frac{1}{4\pi} \int_{-\infty}^{\infty} \text{d}\omega \, \text{Tr} \left[ \ln \left( 1 - V \chi(i\omega) \right) + V \chi(i\omega) \right]
	\label{eq:rpa}
\end{equation}
where $V$ is the Coulomb potential and $\chi$ represents the dynamical charge susceptibility calculated based on $GW$ quasi-particles.

Since the accuracy of the RPA correlation energy sensitively relies on the accuracy of single-particle excitation spectra, we further employ the $GW$ approximation \cite{hedin-gw-pr65,louie-ppa-prb86,aryasetiawan-gw-rpp98,rubio-rmp02,reining-gw-wires18,golze-gw-fc19} to improve the HF single-particle spectra. In the $GW$ approximation, the electron self energy $\Sigma_c$ is calculated as:
\begin{equation}
	\Sigma_c(\br, \br', \omega) = \frac{i}{2\pi} \int \text{d}\nu \, e^{i\nu\delta^+} G_0(\br, \br', \omega + \nu) W_{\text{RPA}}(\br', \br, \nu)
	\label{eq:gw-self-energy}
\end{equation}
where $G_0$ is the HF Green's function and $W_{\text{RPA}}$ is the dynamically screened Coulomb interaction:
\begin{equation}
	W_{\text{RPA}}(\br, \br', \omega) = \int \text{d}\br'' \left[ \epsilon^{-1}_{\text{RPA}}(\br, \br'', \omega) - \delta(\br, \br'') \right] V(\br'', \br')
	\label{eq:w}
\end{equation}
where the static part has been subtracted as it is already taken into account in the HF calculations. In this formalism, the inverse dielectric function $\epsilon^{-1}_{\text{RPA}}$ plays a critical role in capturing the frequency-dependent screening of the Coulomb interaction. The quasiparticle (QP) energies are then corrected through the $GW$ self energy, expressed as:
\begin{equation}
	\varepsilon_{n\bk}^{\text{QP}} = \varepsilon_{n\bk}^{\text{HF}} + Z_{n\bk} \, \text{Re} \, \Sigma_c(\bk, \varepsilon_{n\bk}^{\text{HF}})_{nn}
	\label{eq:qp}
\end{equation}
where $Z_{n\bk}=\left[ 1 - \text{Re} \left( \partial \Sigma_c(\bk, \omega)_{nn}/\partial \omega \right)_{\omega = \varepsilon_{n\bk}^{\text{HF}}} \right]^{-1}$ is the QP weight, accounting for interaction renormalization effects of QPs in the FL state.
We employ the multiple plasmon pole approximation (MPA) \cite{mpa-prb21,mpa-prb23,supp_info} to model the frequency dependent dielectric function, which allows for a  precise and efficient treatment of the screening effects, as discussed with greater details in Supplementary Information \cite{supp_info}. 
These methods allow us to obtain a more accurate and balanced description of both exchange and correlation effects. In the following, we will first benchmark the $GW$+RPA method using the WC transition problem in conventional 2DEG. Then, we will apply this method to slightly charge-doped RMG and discuss the Wigner-crystal transitions in these systems. In our calculations, including both Hartree-Fock and $GW$+RPA calculations, a $9 \times 9$ mesh of reciprocal lattice points has been used for the 2DEG system, while a $7 \times 7$ mesh is used for RMG systems. The mini Brillouin zone (of the presumable WC state) is sampled by an $18 \times 18$ $\bk$ mesh.

\section{$GW$+RPA method benchmarked using conventional 2DEG}
\label{sec:2deg}
We first apply the $GW$+RPA method to conventional 2DEG system to study its Wigner crystallization problem, which yields significantly improved results compared with HF and HF+RPA calculations. 
The 2DEG system, with its kinetic energy expressed as
\begin{equation}   
	H_{\rm{2DEG}}^0(\bk) = \frac{\hbar^2 \bk^2}{2m^*}\,,
\end{equation}
serves as a fundamental platform for investigating many-body physics, including the Wigner crystallization and its transition to FL state. In our work, the $e$-$e$ interaction in 2DEG is described by a dual-gate screened Coulomb interaction as expressed in Eq.~\eqref{eq:coulomb}.
The key parameter controlling the WC transition is the dimensionless Wigner-Seitz radius $r_s=g_\nu g_s m^*/(\sqrt{\pi n} \epsilon_r m_0 a_B)$, where$g_\nu$ and $g_s$ are the (possible) valley and spin degeneracies, respectively, $n$ is the electron density, $\epsilon_r$ is the relative dielectric constant, $m^*$ is the effective mass, $m_0$ is the electron mass, and $a_B$ is Bohr radius.  The Wigner-Seitz radius measures the average inter-electron separation
and characterizes the balance between interaction and kinetic energies.  

\begin{figure*}[htb]
	\centering
	\includegraphics[width=5in]{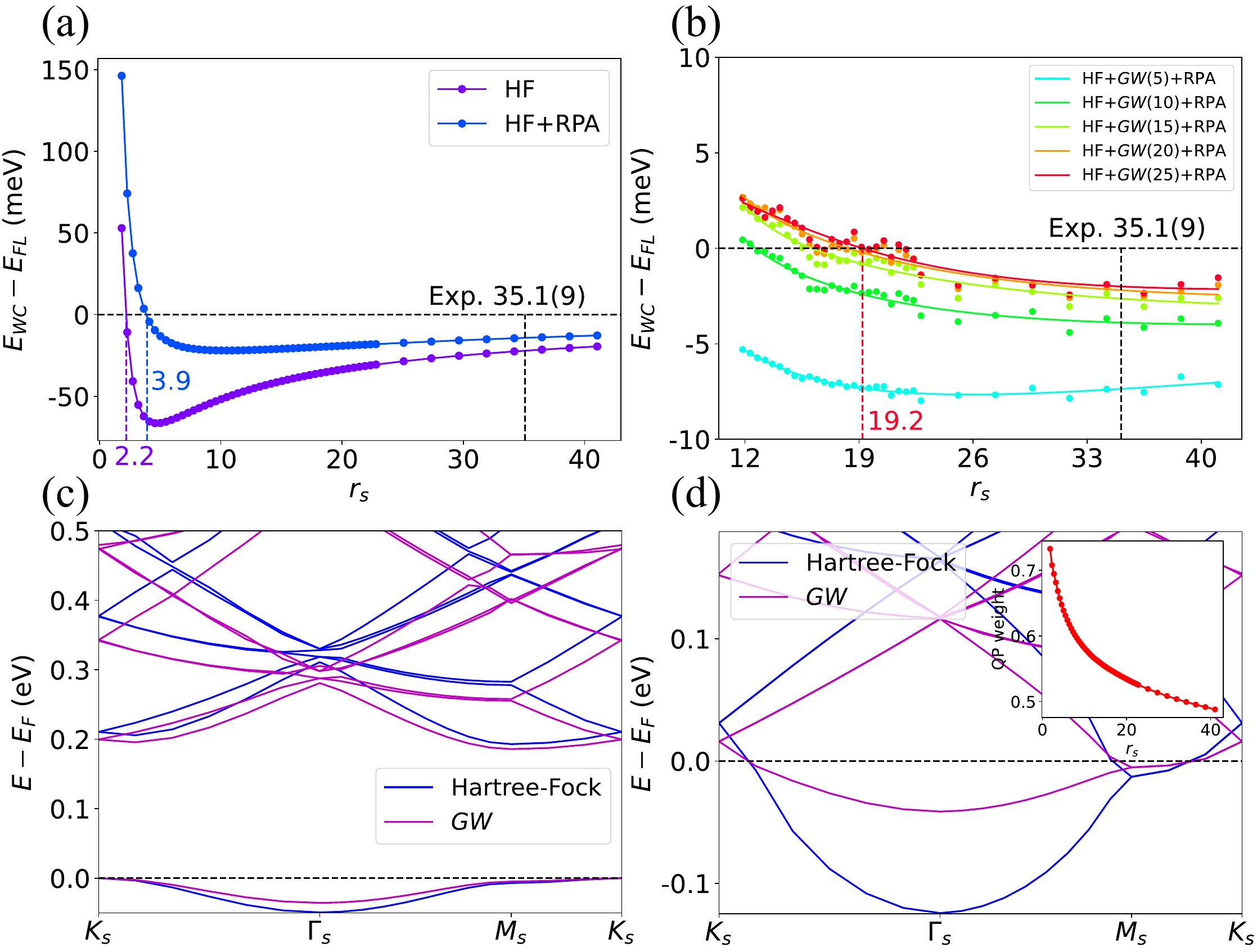}
	\caption{(a) Condensation energy of WC state of conventional 2DEG as a function of $r_s$ based on HF and HF+RPA calculations. (b) Condensation energy as a function of $r_s$ based on HF+$GW$+RPA calculations with increasing number of $GW$ bands. (c) Band structure of fully spin-polarized WC state at $r_s = 11.4$. (d) Band structure of spin-degenerate FL state at $r_s = 11.4$.} 
\label{fig:1}
\end{figure*}

In Fig.~\ref{fig:1}(a), we show the condensation energy,  $E_{\rm{cond.}} = E_{\text{WC}} - E_{\text{FL}}$ as a function of $r_s$, where $E_{\rm{WC}}$ ($E_{\rm{FL}}$) denotes total energy of WC (FL) state. The results from HF calculations, as marked by the purple line, significantly overestimate the stability of the WC state by neglecting electron correlation effects. Under HF approximation, the WC transition occurs at a small critical value $r_s^* \sim 2.2$, much smaller than the experimental value 35.1(9) \cite{wigner-exp-prl99}. The blue line in Fig.~\ref{fig:1}(a) shows the results after incorporating RPA correlation energy, which captures collective charge-fluctuation effects and partly corrects HF's bias towards WC state. Including the RPA correlation energy shifts the WC transition to larger values of $r_s \sim 3.9$, highlighting the critical role of correlations in stabilizing the FL state.

We further refined these calculations using the $GW$+RPA method, as shown in Fig.~\ref{fig:1}(b). The inclusion of additional $GW$ corrections to the HF bands (labeled by ``$GW$(5-25)'' with 5-25 denoting the number of HF bands corrected by $GW$ self energies) progressively shifts the critical $r_s$ to larger values, ultimately converges around $r_s^* \sim 19.2$ (data fitting is described in Supplementary Information \cite{needs-wigner-qmc-prl09,rapisarda-dqmc-1996,supp_info}), a value that matches experiment ($r_s^*\sim 35.1$)  much closer than those without $GW$ corrections. This demonstrates that the combination of  $GW$ corrections and RPA correlation energy better captures the delicate balance between exchange and correlation effects, providing a more accurate prediction of the WC transition.

Fig.~\ref{fig:1}(c) shows the QP band structure of a fully spin-polarized WC state at $r_s = 11.4$. A gap opens at the Brillouin zone (of the Wigner lattice) boundary due to strong Coulomb interactions. When the $GW$ correction is applied, charge fluctuations screen part of the Coulomb interaction, which  reduce both the HF gap and the bandwidth of the WC state. Fig.~\ref{fig:1}(d) presents the band structure of the spin-degenerate FL state at the same $r_s = 11.4$. After applying the $GW$ correction, the bandwidth is significantly decreased, leading to much larger effective mass due to correlation effects. The significant reduction in the bandwidth of the FL state leads to a  much larger absolute value of RPA correlation energy, which counterbalances the HF's bias towards WC state.
Additionally, from the GW calculation we obtain the QP weight for the FL state, as shown in the inset of Fig.~\ref{fig:1}(d). As $r_s$ increases, the QP weight gradually decreases, implying more significant correlation effects.
These results suggest the $GW+$RPA method can faithfully describe the WC transition problem in conventional 2DEG, thus can be readily extended to other  2D systems with long-range Coulomb interactions such as RMG.

\section{Rhombohedral multilayer graphene described by Dirac-fermion models}
After being benchmarked in conventional 2DEG, we continue to apply the $GW$+RPA framework to slightly charge-doped $n$-layer RMG with $n=2, 3, 4, 5, 6$. 
We first start with a $n$-order Dirac fermion model  which approximately characterizes the low-energy non-interacting Hamiltonian of $n$-layer RMG \cite{min-multilayer-08}, expressed as:
\begin{equation}
	H^0_{\mu,n}(\bk) = -
	\begin{pmatrix}
		\Delta & t_\perp(\nu_\mu^\dagger)^n \\ 
		t_\perp(\nu_\mu)^n & -\Delta
	\end{pmatrix}
	\label{eq:H-dirac}
\end{equation}
where $\nu_\mu = \hbar v_F (\mu k_x + ik_y)/t_\perp$, $\hbar v_F=5.25\,\text{eV}\cdot\text{\AA}$ is the in-plane Fermi velocity, and $t_\perp=0.34\,$eV is the nearest interlayer hopping amplitude, $\mu=\mp$ denotes $K/K'$ valley of graphene. $\Delta$ denotes the Dirac-fermion mass  which physically originates from vertical electric field applied to RMG. These parameters are all derived from a realistic Slater-Koster tight-binding model of graphene \cite{moon-tbg-prb13}. For all calculations based on such Dirac-fermion models, we set $\Delta = 0.1\,$eV.  The above model results from a perturbative expansion of $\nu_{\mu}$ \cite{min-multilayer-08}, which neglects further-neighbor interlayer hopping terms. Later we will consider the next-neighbor hopping effects, such as trigonal warpings in bilayer graphene \cite{blg-prl06}, and we will see that the conclusions are qualitatively unchanged.


Before tackling with multilayer graphene, we first study the single-particle spectrum of gapless Dirac fermions in monolayer graphene at 10\% electron doping with respect to charge neutrality. Besides the main Dirac-cone-like bands, our $GW$ calculations suggest that there are also surrounding satellite features \cite{supp_info}. These satellites arise from electron-plasmon couplings, representing plasmaron (or plasmon-polaron) states, consistent with previous theoretical and experimental reports \cite{graphene-arpes-science10,louie-gw-graphene-prl08,zhou-plasmaron-npjqm21}. This further confirms that our theoretical framework can be safely applied to RMG.
Then, we continue to study $n$-layer rhombohedral graphene ($n=2$, 3, 4, 5, and 6) using $GW$+RPA method. Here we assume that at low carrier densities ($\lessapprox 10^{11}\,\textrm{cm}^{-2}$), the ground state is fully spin-valley polarized as observed in experiments \cite{ju-chern-natnano2023,zhou-trilayer-nature21}, so that flavor (spin/valley) fluctuations are expected to make less important contributions to self energy and correlation energy compared to charge fluctuations within a single flavor.  

\subsection{Hartree-Fock results}
\begin{figure*}[htb]
	\centering
	\includegraphics[width=5in]{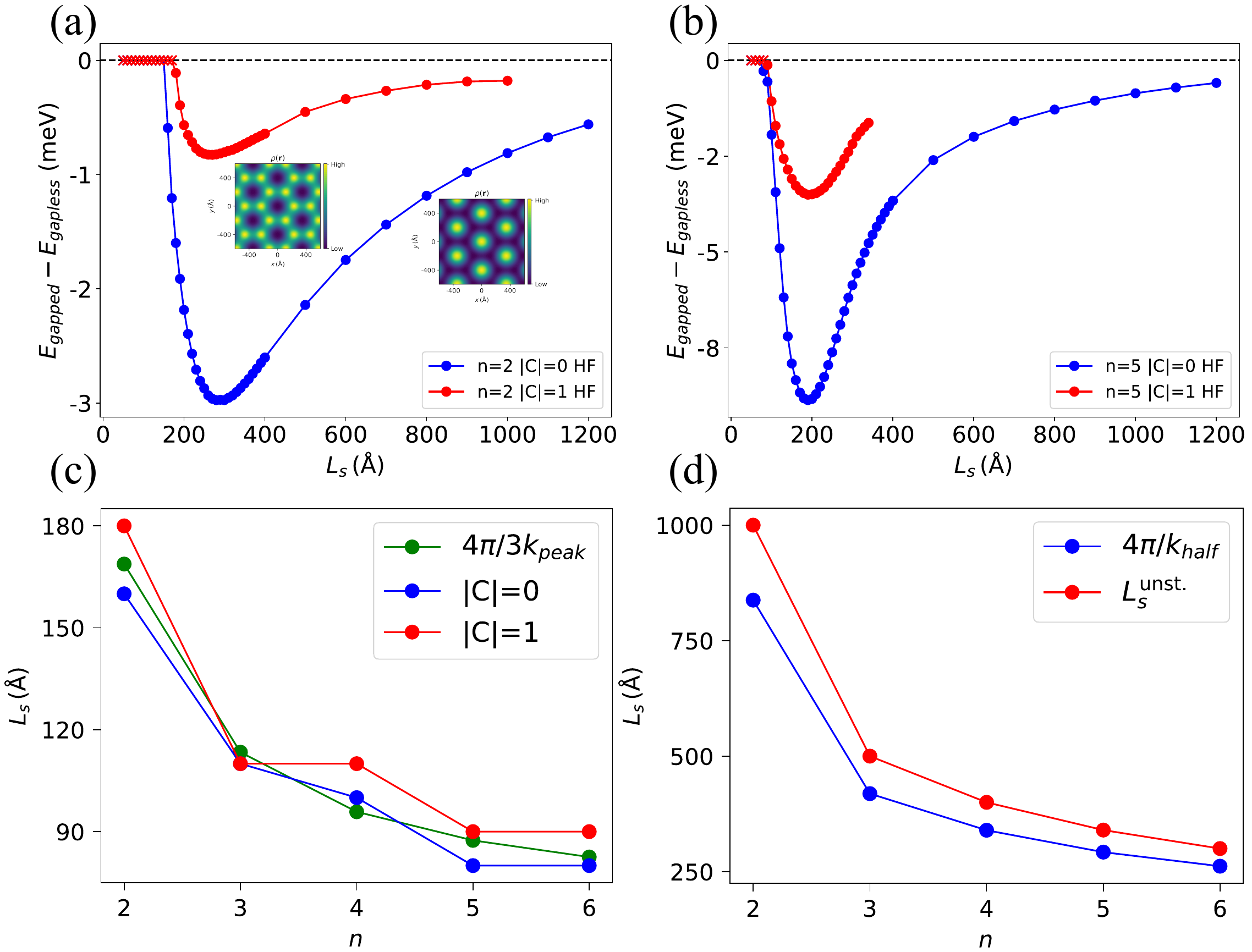}
	\caption{(a) Condensation energies as a function of $L_s$ in bilayer graphene based on HF calculations. The insets show real-space electron distributions for two distinct WC states. (b) Condensation energies as a function of $L_s$ in pentalayer graphene based on HF calculations. (c) $L_s^*$ at which the WC transition onsets as a function of the layer number $n$, calculated using HF approximation. The Green circles represent the values estimated using $4\pi/3 k_{\rm{peak}}$. The blue and red sold circles denote the values obtained from actual numerical HF calculations for $C=0$ and $C=1$ WC states, respectively. (d) $L_s^{\rm{unst}.}$ and $4\pi / k_{\rm{half}}$ as a function of $n$ (see text).}
	\label{fig:3}
\end{figure*}

Fig.\ref{fig:3}(a) shows the WC condensation energy as a function of WC lattice constant $L_s$ for bilayer graphene calculated only in HF approximation. Two topologically distinct WC states are obtained: the topologically trivial Chern-number-0 WC state (blue line) emerges for $L_s \geq 160\,\text{\AA}$, while the topologically nontrivial WC state with Chern-number 1 (red line) is stable within a finite range $180\,\text{\AA}\lessapprox L_s \lessapprox 1000\,\text{\AA}$. The Chern-number-0 WC state consistently has lower energy, which is energetically favorable under HF approximation. The insets in Fig.\ref{fig:3}(a) display the real-space electron distributions: the left inset shows that of the Chern-number-1 WC state, where electrons form a honeycomb lattice \cite{cano-ahc-prl24}, and the right inset shows the Chern-number-0 WC state, where electrons form a triangular lattice. Similarly, Fig.~\ref{fig:3}(b) presents the results for the pentalayer system ($n=5$).
The Chern-number-0 WC state (blue line) onsets for $L_s \geq 80\,\text{\AA}$, while the Chern-number-1 AHC state (red line) emerges in a smaller range $90\,\text{\AA}\lessapprox L_s \lessapprox 340\,\text{\AA}$. Again, the topologically trivial WC state always has lower HF energy. This is because the exchange energy of trivial WC state is larger (in amplitude) than that of the topological AHC state. 
In Supplementary Information we provide the HF band structures of both topologically trivial and non-trivial WC states for $n$-layer RMG from $n=2$ to $6$. A general feature is that the single-particle charge gap of the topological AHC state is much smaller than that of the trivial WC state. Later we will see that this feature remains unchanged even after including $GW$ self-energy corrections (see Fig.~\ref{fig:5}(d)), which eventually makes the AHC state being more energetically stable than the trivial WC state after taking into account correlation effects due to dynamical charge fluctuations.

It is worth noting that the Berry curvature of the non-interacting conduction band of the $n$-order Dirac-fermion model can be calculated analytically,
	$\Omega(\bk,n)=\frac{\gamma^2 n^2 \Delta k^{2n-2}}{2(\gamma^2 k^{2n}+\Delta^2)^{3/2}}$,
where $\gamma=(-1/t_\perp)^{n-1}v_F^n$. The Berry curvature $\Omega(\bk,n)$ is peaked at the wavevector $k_{\rm{peak}}$,
	$k_{\rm{peak}}=\frac{1}{v_F}\,\Big(\frac{n-1}{n+2}\Big)^{\frac{1}{2n}}\,\Delta^{\frac{1}{n}}\,t_{\perp}^{\frac{n-1}{n}}$. 
As the number of layers increases, $k_{\rm{peak}}$ shifts toward larger value.  HF calculations indicate that whenever the Fermi wavevector $k_F\sim 4\pi/3L_s^*$ coincides with $k_{\rm{peak}}$, the system would undergo WC transition with the critical WC lattice constant $L_s^*$.
This means that, at the HF level, WC state would emerge at larger carrier density with increasing number of layers $n$.
Fig~\ref{fig:3}(c) shows the variation of critical lattice constant $L_s^*$ (at the WC transition onsets) as a function of the layer number $n$. It can be seen that the $L_s^*$ estimated using $4\pi/3 k_{\rm{peak}}$ closely matches the actual results extracted from numerical HF calculations.
Moreover, the Chern-number-1 AHC states become unstable at large $L_s$ for all of the $n$-layer systems, because at large $L_s$ the Berry curvature would eventually  be pushed out of the first Brillouin zone (of the corresponding WC) completely, thus it is impossible to yield a Chern-number-1 state. Fig~\ref{fig:3}(d) presents the calculated lattice constant at which the Chern-number-1 WC state becomes unstable (denoted by $L_s^{\rm{unst.}}$) as a function of $n$, which is denoted by red filled circles. In the same figure, we also plot $4\pi$ divided by the wavevector marking the half-peak-value of the Berry curvature (denoted as $k_{\rm{half}}$) using solid blue circles. It can be seen that $L_s^{\rm{unst}.}$ and $4\pi / k_{\rm{half}}$ almost follow exactly the same trend as $n$ increases. More detailed results of the HF calculations of  WC transition in the $n$-layer systems are provided in the Supplementary Information \cite{supp_info}.

\subsection{$GW$+RPA results}
\begin{figure*}[htb]
\centering
\includegraphics[width=5.5in]{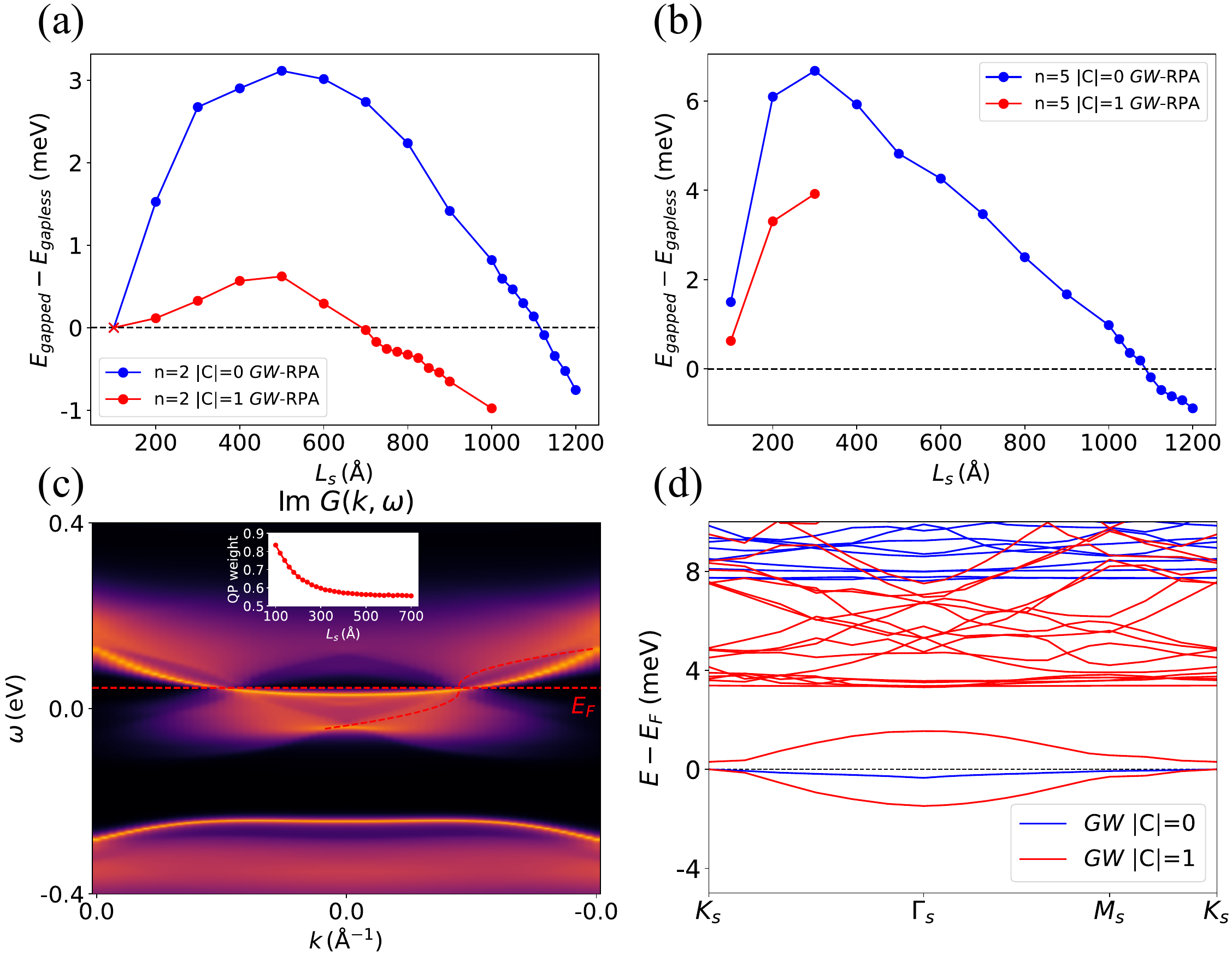}
\caption{(a) Condensation energies as a function of $L_s$  based on  $GW$+RPA calculations, (a) for bilayer graphene, and (b) for pentalayer graphene. (c) Single-particle spectrum of the FL state for $n=2$ at $L_s = 200\,\text{\AA}$. The inset shows the QP weight for the FL state as a function of $L_s$. (d)  $GW$ quasiparticle band structures of the Chern-number-0 and the Chern-number-1 WC states for $n=2$ at $L_s = 800\,\text{\AA}$.}
\label{fig:4}
\end{figure*}

Figs.\ref{fig:4}(a) and \ref{fig:4}(b) illustrate the effects of including RPA correlation energy based on $GW$-corrected quasi-particle bands for    bilayer system  and  pentalayer system, respectively. The blue and red lines denote the WC condensation energies of the Chern-number-0 and the Chern-number-1 WC states. In both systems, the inclusion of correlation effects through $GW$+RPA approach would significantly mitigate the exchange energy bias towards the WC state, shifting the critical lattice constant of the topologically trivial WC phase (denoted by $L_s^*$) from $L_s^*\sim200\,\text{\AA}$ to a much larger value $L_s^*\sim1100\,\text{\AA}$, corresponding to a critical density $\sim 1\times 10^{10}\,\rm{cm}^{-2}$.  Our calculations seem to suggest this critical lattice constant seems to be a  universal one for all $n$-layer Dirac-fermion models ($n=2, 3, 4, 5, 6$) with fixed model parameters.  

Notably, the correlation effects  further lower the total energy of the Chern-number-1 WC (also known AHC) state below that of the Chern-number-0 state, particularly for the bilayer system. As clearly shown in Fig.~\ref{fig:4}(a),  for bilayer Dirac-fermion model, when $L_s$ is greater than a critical value of $700\,\text{\AA}$ (corresponding to a density $\sim 2.4\times 10^{10}\,\rm{cm}^{-2}$), the AHC becomes the ground state over both  FL and  trivial WC. This is because the energy gap of the AHC state ($C=1$ state) is much smaller than that of the  trivial WC ($C=0$ state), as clearly shown in Fig.~\ref{fig:4}(d). This results in larger charge susceptibility and  a lower  RPA correlation energy (see Eq.~\eqref{eq:rpa}) for the AHC state, which ultimately allows it to be the ground state over the topologically trivial WC for the bilayer system. In contrast, in the cases of $n\geq 3$, the AHC state becomes less stable \cite{supp_info} because the Berry curvature is peaked at larger wavevector for larger $n$, and would be completely pushed out of the first Brillouin zone (of the presumable WC) for large $L_s$. As a result, our $GW$+RPA calculations suggest that AHC cannot be the actual ground state of the $n$-order Dirac-fermion models for $n\geq 3$. For example, the WC condensation energy for pentalayer system ($n=5$) is shown in Fig.~\ref{fig:4}(b). It is clearly seen that AHC is only a metastable state for $L_s\lessapprox 340\,\text{\AA}$, but always has higher energy than the FL state.  Similar conclusion holds  for $n= 3, 4, 6$ \cite{supp_info}.

Fig.\ref{fig:4}(c) further shows the  $GW$ single-particle spectrum of the FL state for the bilayer system  at $L_s = 200\,\text{\AA}$, which reveals plasmon satellites due to strong electron-plasmon couplings, similar to those observed in monolayer graphene. A detailed discussion of these features is provided in  Supplementary Information \cite{supp_info}. The inset in Fig.~\ref{fig:4}(c) shows how the QP weight decreases with the increase of $L_s$, suggesting stronger correlation effects at lower density.

\section{Trigonal warping effects in BLG}

\begin{figure*}[htb]
\centering
\includegraphics[width=6.5in]{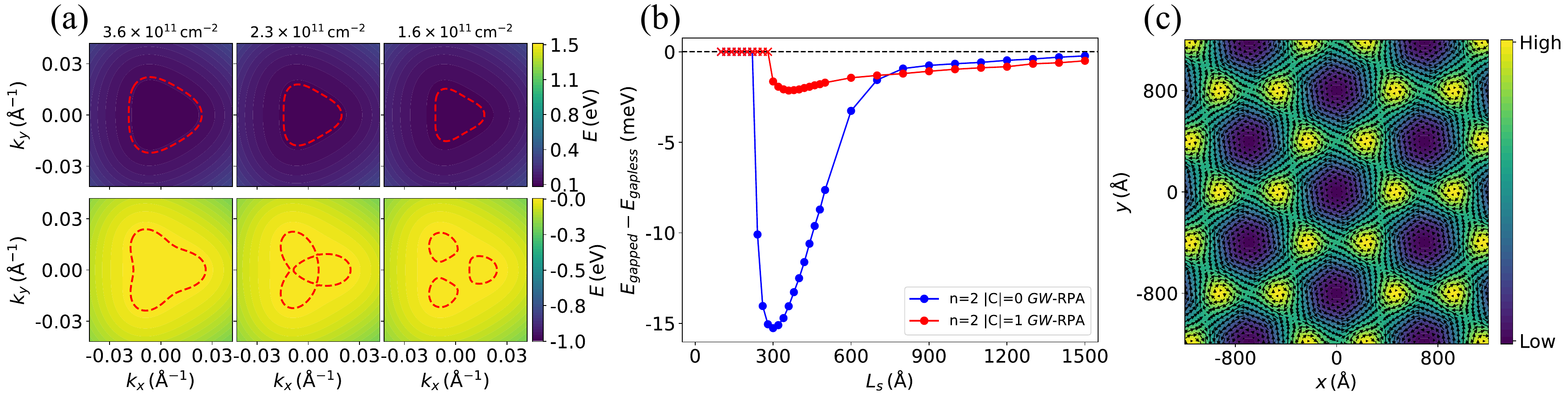}
\caption{(a) Evolution of non-interacting Fermi surfaces of electron-doped (upper panel) and hole-doped (lower panel) bilayer graphene including trigonal warping effects. (b) $GW$+RPA condensation energies of both trivial WC (blue dots) and AHC (red dots) in electron-doped BLG as a function of $L_s$, including trigonal warping effects. (a) Ground-state current-density distribution (black arrows) of AHC in electron-doped BLG at $L_s = 800\,\text{\AA}$. The color coding indicates the absolute value of charge density.}
\label{fig:5}
\end{figure*}

To model a realistic bilayer graphene system more accurately, we continue to include the  next-neighbor interlayer hopping terms expanded to the leading order of $\mathbf{k}$, then downfold the full bilayer Hamiltonian to the low-energy Hilbert space consisted of the two states from the $A$ sublattice of the bottom layer and $B$ sublattice of the top layer. The inclusion of further-neighbor interlayer hopping would result in a Hamiltonian $H^{\rm{tri.}}_{\mu,2}$ with extra terms from trigonal warping \cite{blg-prl06}, which is expressed as 
\begin{widetext}
	\begin{equation}
		H^{\text{tri.}}_{\mu,2}(\mathbf{k}) = 
		\begin{pmatrix}
			-\Delta + \frac{2 v_F t_{\perp}+(v_F^2-v_\perp^2)\Delta}{t_{\perp}^2+\Delta^2} p_\mu p_\mu^\dagger& v_{\perp} p_\mu-\frac{(v_F^2+v_\perp^2) t_{\perp}}{t_{\perp}^2+\Delta^2}(p_\mu^\dagger)^2 \\ 
			v_{\perp} p_\mu^\dagger-\frac{(v_F^2+v_\perp^2) t_{\perp}}{t_{\perp}^2+\Delta^2}(p_\mu)^2 & \Delta + \frac{2 v_F t_{\perp}-(v_F^2-v_\perp^2)\Delta}{t_{\perp}^2+\Delta^2} p_\mu p_\mu^\dagger
		\end{pmatrix},
	\end{equation}
\end{widetext}
where $p_\mu = \hbar (\mu k_x + ik_y)$, and $\hbar v_{\perp}=0.447\,\text{eV}\cdot\text{\AA}$ is the leading-order coefficient of $\mathbf{k}$ in the next-neighbor interlayer hopping. We set $\Delta=0.05\,$eV for all calculations including trigonal warping effects of bilayer graphene, corresponding to an experimentally accessible displacement field $\sim 1\,$V/nm.
 This term will break the continuous rotational symmetry of the Dirac-fermion model, leading to trigonal warping of the band structure. It also breaks the particle-hole symmetry of the $n$-order Dirac-fermion model described by Eq.~\eqref{eq:H-dirac}. Correspondingly, the Berry curvature is no longer peaked along an isotropic circle at $k_{\rm{peak}}$, rather becomes more anisotropic with discrete threefold symmetry as shown in the right panel of Fig.~\ref{fig:0}(b). 
Moreover, due to the breaking of  particle-hole symmetry, non-interacting Fermi surfaces on the electron-doped side and hole-doped side becomes dramatically different. Indeed, previous transport measurements suggest distinct superconducting behaviors when BLG is in proximity with WSe$_2$ \cite{li-blg-nature24}. As shown in the lower panel of Fig.~\ref{fig:5}(a), on the hole-doped side, there is a single Fermi surface when the carrier density $\rho=3.6\times 10^{11}\,\text{cm}^{-2}$, which evolves to a trinity-knot shaped Fermi surface at $\rho=2.3\times 10^{11}\,\rm{cm}^{-2}$, then eventually splits to three degenerate small Fermi pockets when $\rho=1.6\times 10^{11}\,\rm{cm}^{-2}$.  In contrast, as shown in the upper panel of Fig.~\ref{fig:5}(a), there is always a single Fermi surface with three-fold anisotropy on the electron-doped side. This suggests that the interacting ground states on the electron-doped and hole-doped sides may be completely different. Previous works suggest the presence of diverse correlated states on the hole-doped side including spin-valley-polarized metal \cite{young-blg-science22,weitz-blg-nature22} and magnetic-field-stabilized superconductivity \cite{young-blg-science22} etc. 
On the electron-doped side, however, the system has simpler  Fermi-surface topology, which bears more resemblance to  the Dirac-fermion model. Thus we may expect the emergence of AHC on the electron-doped side.
Then, we continue to study the interacting ground state of slightly electron-doped bilayer graphene using $GW$+RPA approach including trigonal warping effects.

The blue and red dots  in Fig.\ref{fig:5}(b) shows the $GW$+RPA condensation energies of the trivial WC state and AHC state including trigonal-warping effects, respectively. We  see that the critical lattice constant $L_s^*$ (below which the condensation energy is negative) is shifted to much higher value compared to the case of Dirac-fermion model (see Fig.~\ref{fig:4}(a)), with $L_s^*\approx 300\,\text{\AA}$ for AHC, corresponding to a critical density $\sim 1.3\times 10^{11}\,\textrm{cm}^{-2}$, and $L_s^*\approx 240\,\text{\AA}$ for trivial WC, corresponding to a critical density $\sim 2.0\times 10^{11}\,\textrm{cm}^{-2}$. Most saliently,  AHC becomes the ground state over both trivial WC  and FL  when $L_s\gtrapprox 750\,\AA$ with $\rho\lessapprox 2\times 10^{10}\,\rm{cm}^{-2}$, even after taking into account the trigonal warping effects. Again, the AHC state is energetically stabilized over the trivial WC due to the lower correlation energy gained from dynamical charge fluctuations.

We continue to study the ground-state properties of AHC. Different from trivial WC which spontaneously forms a triangular lattice in real space, the charge density distribution of AHC forms an emergent honeycomb lattice in real space \cite{cano-ahc-prl24}. Moreover, since AHC spontaneously breaks time-reversal symmetry, the ground state carries spontaneous current density as shown by the black arrows in Fig.~\ref{fig:5}(c). The ground-state currents wind around the charge-density peaks in clockwise manner, while they wind around the central region of charge-density minima in a counter-clockwise way, with the maximal current density $\sim 0.055\,\rm{nA}/\AA^{2}$.  Such ground-state current pattern usually corresponds to a large orbital magnetization as in the case of twisted graphene systems \cite{jpliu-prx19}. As a result, the AHC may be further stabilized by vertical magnetic field due to large orbital Zeeman couplings.

\section{Discussions}
\label{sec:discuss}

To summarize, we have developed a $GW$+RPA framework to study the WC transitions in generic interacting 2D systems at low carrier densities. After being benchmarked using conventional 2DEG system, we adopt this framework to study the ground-state properties and single-particle excitations in slightly charge-doped RMG. Mean-field calculations within HF approximation indicate that both topologically nontrivial AHC and trivial WC states can emerge  for all  the $n$-layer RMG ($n=2, 3, 4, 5, 6$), with the condensation energy of trivial WC being consistently lower than that of AHC.  However, dynamical charge fluctuation effects as faithfully captured by $GW$+RPA method, can significantly de-stabilize trivial WC state, shifting the WC transition point to a much lower critical density $\sim  10^{10}\,\rm{cm}^{-2}$. Most saliently, using a realistic modeling of bilayer graphene including trigonal warping effects, we find that the topological AHC state emerges in electron-doped BLG when the carrier density is below $\sim 1.3\times 10^{11}\,\text{cm}^{-2}$, and  becomes the unique ground state over both the trivial WC and FL when the  density is lower than $\sim 2\times 10^{10}\,\textrm{cm}^{-2}$. Such AHC state in bilayer graphene is stabilized due to a subtle interplay between exchange and correlation effects: on the one hand,  the exchange energy of the charge-gapped, symmetry-breaking AHC state is much lower than that of the gapless, symmetry-preserving FL state, which makes the total energy of AHC being lower than that of FL state at sufficiently low carrier densities; on the other hand, the correlation energy of AHC turns out to be  lower than that of the trivial WC state at low densities due to the much smaller single-particle charge gap (see Fig.~\ref{fig:4}(d)) thus stronger charge-fluctuation effects, which eventually helps AHC to be the genuine ground state over both FL and trivial WC.  The AHC state forms an emergent honeycomb lattice, and carries an intriguing ground-state current pattern in real space.  In contrast, when the number of layers $n\geq 3$, based on the simplified $n$-order Dirac-fermion model, we find that the Berry curvatures of non-interacting Bloch functions are distributed at wavevectors that are further away from the central Dirac points, thus can be easily pushed out of the first Brillouin zone (of the presumable WC) at low densities. As a result, it may be harder to stabilize the topological AHC state when the number number of layers is greater than 2.  
Therefore, we propose that bilayer graphene under slight carrier doping is one of the most promising candidates to realize the AHC state.

The understanding of  AHC state is still in a preliminary stage. A lot of its physical properties, such as the collective excitations, response to electromagnetic fields, and quantum critical behavior near phase transitions, are still open questions.  Our work may stimulate further experimental and theoretical explorations of the  AHC state, with more focus on the pristine bilayer graphene system. 
Moreover, the theoretical framework developed in this work can be readily applied to other interacting 2D systems including moir\'e superlattices.




\section{Acknowledgements}
We thank Xinguo Ren, Yves Hon Kwan, and Bogdan A. Bernevig for valuable discussions. This work is supported by the National Key R \& D program of China (grant No. 2024YFA1410400 and grant No. 2020YFA0309601) and the National Natural Science Foundation of China (grant No. 12174257).

\bibliography{tmg_tidy}

\begin{thebibliography}{51}%
\makeatletter
\providecommand \@ifxundefined [1]{%
 \@ifx{#1\undefined}
}%
\providecommand \@ifnum [1]{%
 \ifnum #1\expandafter \@firstoftwo
 \else \expandafter \@secondoftwo
 \fi
}%
\providecommand \@ifx [1]{%
 \ifx #1\expandafter \@firstoftwo
 \else \expandafter \@secondoftwo
 \fi
}%
\providecommand \natexlab [1]{#1}%
\providecommand \enquote  [1]{``#1''}%
\providecommand \bibnamefont  [1]{#1}%
\providecommand \bibfnamefont [1]{#1}%
\providecommand \citenamefont [1]{#1}%
\providecommand \href@noop [0]{\@secondoftwo}%
\providecommand \href [0]{\begingroup \@sanitize@url \@href}%
\providecommand \@href[1]{\@@startlink{#1}\@@href}%
\providecommand \@@href[1]{\endgroup#1\@@endlink}%
\providecommand \@sanitize@url [0]{\catcode `\\12\catcode `\$12\catcode
  `\&12\catcode `\#12\catcode `\^12\catcode `\_12\catcode `\%12\relax}%
\providecommand \@@startlink[1]{}%
\providecommand \@@endlink[0]{}%
\providecommand \url  [0]{\begingroup\@sanitize@url \@url }%
\providecommand \@url [1]{\endgroup\@href {#1}{\urlprefix }}%
\providecommand \urlprefix  [0]{URL }%
\providecommand \Eprint [0]{\href }%
\providecommand \doibase [0]{https://doi.org/}%
\providecommand \selectlanguage [0]{\@gobble}%
\providecommand \bibinfo  [0]{\@secondoftwo}%
\providecommand \bibfield  [0]{\@secondoftwo}%
\providecommand \translation [1]{[#1]}%
\providecommand \BibitemOpen [0]{}%
\providecommand \bibitemStop [0]{}%
\providecommand \bibitemNoStop [0]{.\EOS\space}%
\providecommand \EOS [0]{\spacefactor3000\relax}%
\providecommand \BibitemShut  [1]{\csname bibitem#1\endcsname}%
\let\auto@bib@innerbib\@empty
\bibitem [{\citenamefont {Wigner}(1934)}]{wigner-pr34}%
  \BibitemOpen
  \bibfield  {author} {\bibinfo {author} {\bibfnamefont {E.}~\bibnamefont
  {Wigner}},\ }\href {https://doi.org/10.1103/PhysRev.46.1002} {\bibfield
  {journal} {\bibinfo  {journal} {Phys. Rev.}\ }\textbf {\bibinfo {volume}
  {46}},\ \bibinfo {pages} {1002} (\bibinfo {year} {1934})}\BibitemShut
  {NoStop}%
\bibitem [{\citenamefont {Andrei}\ \emph {et~al.}(1988)\citenamefont {Andrei},
  \citenamefont {Deville}, \citenamefont {Glattli}, \citenamefont {Williams},
  \citenamefont {Paris},\ and\ \citenamefont {Etienne}}]{eva-wigner-prl88}%
  \BibitemOpen
  \bibfield  {author} {\bibinfo {author} {\bibfnamefont {E.~Y.}\ \bibnamefont
  {Andrei}}, \bibinfo {author} {\bibfnamefont {G.}~\bibnamefont {Deville}},
  \bibinfo {author} {\bibfnamefont {D.~C.}\ \bibnamefont {Glattli}}, \bibinfo
  {author} {\bibfnamefont {F.~I.~B.}\ \bibnamefont {Williams}}, \bibinfo
  {author} {\bibfnamefont {E.}~\bibnamefont {Paris}},\ and\ \bibinfo {author}
  {\bibfnamefont {B.}~\bibnamefont {Etienne}},\ }\href
  {https://doi.org/10.1103/PhysRevLett.60.2765} {\bibfield  {journal} {\bibinfo
   {journal} {Phys. Rev. Lett.}\ }\textbf {\bibinfo {volume} {60}},\ \bibinfo
  {pages} {2765} (\bibinfo {year} {1988})}\BibitemShut {NoStop}%
\bibitem [{\citenamefont {Yoon}\ \emph {et~al.}(1999)\citenamefont {Yoon},
  \citenamefont {Li}, \citenamefont {Shahar}, \citenamefont {Tsui},\ and\
  \citenamefont {Shayegan}}]{wigner-exp-prl99}%
  \BibitemOpen
  \bibfield  {author} {\bibinfo {author} {\bibfnamefont {J.}~\bibnamefont
  {Yoon}}, \bibinfo {author} {\bibfnamefont {C.~C.}\ \bibnamefont {Li}},
  \bibinfo {author} {\bibfnamefont {D.}~\bibnamefont {Shahar}}, \bibinfo
  {author} {\bibfnamefont {D.~C.}\ \bibnamefont {Tsui}},\ and\ \bibinfo
  {author} {\bibfnamefont {M.}~\bibnamefont {Shayegan}},\ }\href
  {https://doi.org/10.1103/PhysRevLett.82.1744} {\bibfield  {journal} {\bibinfo
   {journal} {Phys. Rev. Lett.}\ }\textbf {\bibinfo {volume} {82}},\ \bibinfo
  {pages} {1744} (\bibinfo {year} {1999})}\BibitemShut {NoStop}%
\bibitem [{\citenamefont {Trail}\ \emph {et~al.}(2003)\citenamefont {Trail},
  \citenamefont {Towler},\ and\ \citenamefont {Needs}}]{wc-hf-prb03}%
  \BibitemOpen
  \bibfield  {author} {\bibinfo {author} {\bibfnamefont {J.~R.}\ \bibnamefont
  {Trail}}, \bibinfo {author} {\bibfnamefont {M.~D.}\ \bibnamefont {Towler}},\
  and\ \bibinfo {author} {\bibfnamefont {R.~J.}\ \bibnamefont {Needs}},\ }\href
  {https://doi.org/10.1103/PhysRevB.68.045107} {\bibfield  {journal} {\bibinfo
  {journal} {Phys. Rev. B}\ }\textbf {\bibinfo {volume} {68}},\ \bibinfo
  {pages} {045107} (\bibinfo {year} {2003})}\BibitemShut {NoStop}%
\bibitem [{\citenamefont {Drummond}\ and\ \citenamefont
  {Needs}(2009)}]{needs-wigner-qmc-prl09}%
  \BibitemOpen
  \bibfield  {author} {\bibinfo {author} {\bibfnamefont {N.~D.}\ \bibnamefont
  {Drummond}}\ and\ \bibinfo {author} {\bibfnamefont {R.~J.}\ \bibnamefont
  {Needs}},\ }\href {https://doi.org/10.1103/PhysRevLett.102.126402} {\bibfield
   {journal} {\bibinfo  {journal} {Phys. Rev. Lett.}\ }\textbf {\bibinfo
  {volume} {102}},\ \bibinfo {pages} {126402} (\bibinfo {year}
  {2009})}\BibitemShut {NoStop}%
\bibitem [{\citenamefont {Padhi}\ \emph {et~al.}(2018)\citenamefont {Padhi},
  \citenamefont {Setty},\ and\ \citenamefont
  {Phillips}}]{philips-nanoletter-18}%
  \BibitemOpen
  \bibfield  {author} {\bibinfo {author} {\bibfnamefont {B.}~\bibnamefont
  {Padhi}}, \bibinfo {author} {\bibfnamefont {C.}~\bibnamefont {Setty}},\ and\
  \bibinfo {author} {\bibfnamefont {P.~W.}\ \bibnamefont {Phillips}},\ }\href
  {https://doi.org/10.1021/acs.nanolett.8b02033} {\bibfield  {journal}
  {\bibinfo  {journal} {Nano Lett.}\ }\textbf {\bibinfo {volume} {18}},\
  \bibinfo {pages} {6175} (\bibinfo {year} {2018})}\BibitemShut {NoStop}%
\bibitem [{\citenamefont {Regan}\ \emph
  {et~al.}(2020{\natexlab{a}})\citenamefont {Regan}, \citenamefont {Wang},
  \citenamefont {Jin}, \citenamefont {Utama}, \citenamefont {Gao},
  \citenamefont {Wei}, \citenamefont {Zhao}, \citenamefont {Zhao},
  \citenamefont {Zhang}, \citenamefont {Yumigeta} \emph
  {et~al.}}]{wang-wigner-nature20}%
  \BibitemOpen
  \bibfield  {author} {\bibinfo {author} {\bibfnamefont {E.~C.}\ \bibnamefont
  {Regan}}, \bibinfo {author} {\bibfnamefont {D.}~\bibnamefont {Wang}},
  \bibinfo {author} {\bibfnamefont {C.}~\bibnamefont {Jin}}, \bibinfo {author}
  {\bibfnamefont {M.~I.~B.}\ \bibnamefont {Utama}}, \bibinfo {author}
  {\bibfnamefont {B.}~\bibnamefont {Gao}}, \bibinfo {author} {\bibfnamefont
  {X.}~\bibnamefont {Wei}}, \bibinfo {author} {\bibfnamefont {S.}~\bibnamefont
  {Zhao}}, \bibinfo {author} {\bibfnamefont {W.}~\bibnamefont {Zhao}}, \bibinfo
  {author} {\bibfnamefont {Z.}~\bibnamefont {Zhang}}, \bibinfo {author}
  {\bibfnamefont {K.}~\bibnamefont {Yumigeta}}, \emph {et~al.},\ }\href@noop {}
  {\bibfield  {journal} {\bibinfo  {journal} {Nature}\ }\textbf {\bibinfo
  {volume} {579}},\ \bibinfo {pages} {359} (\bibinfo {year}
  {2020}{\natexlab{a}})}\BibitemShut {NoStop}%
\bibitem [{\citenamefont {Regan}\ \emph
  {et~al.}(2020{\natexlab{b}})\citenamefont {Regan}, \citenamefont {Wang},
  \citenamefont {Jin}, \citenamefont {Bakti~Utama}, \citenamefont {Gao},
  \citenamefont {Wei}, \citenamefont {Zhao}, \citenamefont {Zhao},
  \citenamefont {Zhang}, \citenamefont {Yumigeta}, \citenamefont {Blei},
  \citenamefont {Carlstr{\"o}m}, \citenamefont {Watanabe}, \citenamefont
  {Taniguchi}, \citenamefont {Tongay}, \citenamefont {Crommie}, \citenamefont
  {Zettl},\ and\ \citenamefont {Wang}}]{Regan2020}%
  \BibitemOpen
  \bibfield  {author} {\bibinfo {author} {\bibfnamefont {E.~C.}\ \bibnamefont
  {Regan}}, \bibinfo {author} {\bibfnamefont {D.}~\bibnamefont {Wang}},
  \bibinfo {author} {\bibfnamefont {C.}~\bibnamefont {Jin}}, \bibinfo {author}
  {\bibfnamefont {M.~I.}\ \bibnamefont {Bakti~Utama}}, \bibinfo {author}
  {\bibfnamefont {B.}~\bibnamefont {Gao}}, \bibinfo {author} {\bibfnamefont
  {X.}~\bibnamefont {Wei}}, \bibinfo {author} {\bibfnamefont {S.}~\bibnamefont
  {Zhao}}, \bibinfo {author} {\bibfnamefont {W.}~\bibnamefont {Zhao}}, \bibinfo
  {author} {\bibfnamefont {Z.}~\bibnamefont {Zhang}}, \bibinfo {author}
  {\bibfnamefont {K.}~\bibnamefont {Yumigeta}}, \bibinfo {author}
  {\bibfnamefont {M.}~\bibnamefont {Blei}}, \bibinfo {author} {\bibfnamefont
  {J.~D.}\ \bibnamefont {Carlstr{\"o}m}}, \bibinfo {author} {\bibfnamefont
  {K.}~\bibnamefont {Watanabe}}, \bibinfo {author} {\bibfnamefont
  {T.}~\bibnamefont {Taniguchi}}, \bibinfo {author} {\bibfnamefont
  {S.}~\bibnamefont {Tongay}}, \bibinfo {author} {\bibfnamefont
  {M.}~\bibnamefont {Crommie}}, \bibinfo {author} {\bibfnamefont
  {A.}~\bibnamefont {Zettl}},\ and\ \bibinfo {author} {\bibfnamefont
  {F.}~\bibnamefont {Wang}},\ }\href
  {https://doi.org/10.1038/s41586-020-2092-4} {\bibfield  {journal} {\bibinfo
  {journal} {Nature}\ }\textbf {\bibinfo {volume} {579}},\ \bibinfo {pages}
  {359} (\bibinfo {year} {2020}{\natexlab{b}})}\BibitemShut {NoStop}%
\bibitem [{\citenamefont {Li}\ \emph {et~al.}(2021)\citenamefont {Li},
  \citenamefont {Li}, \citenamefont {Regan}, \citenamefont {Wang},
  \citenamefont {Zhao}, \citenamefont {Kahn}, \citenamefont {Yumigeta},
  \citenamefont {Blei}, \citenamefont {Taniguchi}, \citenamefont {Watanabe},
  \citenamefont {Tongay}, \citenamefont {Zettl}, \citenamefont {Crommie},\ and\
  \citenamefont {Wang}}]{li-wigner-nature21}%
  \BibitemOpen
  \bibfield  {author} {\bibinfo {author} {\bibfnamefont {H.}~\bibnamefont
  {Li}}, \bibinfo {author} {\bibfnamefont {S.}~\bibnamefont {Li}}, \bibinfo
  {author} {\bibfnamefont {E.~C.}\ \bibnamefont {Regan}}, \bibinfo {author}
  {\bibfnamefont {D.}~\bibnamefont {Wang}}, \bibinfo {author} {\bibfnamefont
  {W.}~\bibnamefont {Zhao}}, \bibinfo {author} {\bibfnamefont {S.}~\bibnamefont
  {Kahn}}, \bibinfo {author} {\bibfnamefont {K.}~\bibnamefont {Yumigeta}},
  \bibinfo {author} {\bibfnamefont {M.}~\bibnamefont {Blei}}, \bibinfo {author}
  {\bibfnamefont {T.}~\bibnamefont {Taniguchi}}, \bibinfo {author}
  {\bibfnamefont {K.}~\bibnamefont {Watanabe}}, \bibinfo {author}
  {\bibfnamefont {S.}~\bibnamefont {Tongay}}, \bibinfo {author} {\bibfnamefont
  {A.}~\bibnamefont {Zettl}}, \bibinfo {author} {\bibfnamefont {M.~F.}\
  \bibnamefont {Crommie}},\ and\ \bibinfo {author} {\bibfnamefont
  {F.}~\bibnamefont {Wang}},\ }\href
  {https://doi.org/10.1038/s41586-021-03874-9} {\bibfield  {journal} {\bibinfo
  {journal} {Nature}\ }\textbf {\bibinfo {volume} {597}},\ \bibinfo {pages}
  {650} (\bibinfo {year} {2021})}\BibitemShut {NoStop}%
\bibitem [{\citenamefont {Zhou}\ \emph
  {et~al.}(2021{\natexlab{a}})\citenamefont {Zhou}, \citenamefont {Sung},
  \citenamefont {Brutschea}, \citenamefont {Esterlis}, \citenamefont {Wang},
  \citenamefont {Scuri}, \citenamefont {Gelly}, \citenamefont {Heo},
  \citenamefont {Taniguchi}, \citenamefont {Watanabe} \emph
  {et~al.}}]{mose2-wc-nature21}%
  \BibitemOpen
  \bibfield  {author} {\bibinfo {author} {\bibfnamefont {Y.}~\bibnamefont
  {Zhou}}, \bibinfo {author} {\bibfnamefont {J.}~\bibnamefont {Sung}}, \bibinfo
  {author} {\bibfnamefont {E.}~\bibnamefont {Brutschea}}, \bibinfo {author}
  {\bibfnamefont {I.}~\bibnamefont {Esterlis}}, \bibinfo {author}
  {\bibfnamefont {Y.}~\bibnamefont {Wang}}, \bibinfo {author} {\bibfnamefont
  {G.}~\bibnamefont {Scuri}}, \bibinfo {author} {\bibfnamefont {R.~J.}\
  \bibnamefont {Gelly}}, \bibinfo {author} {\bibfnamefont {H.}~\bibnamefont
  {Heo}}, \bibinfo {author} {\bibfnamefont {T.}~\bibnamefont {Taniguchi}},
  \bibinfo {author} {\bibfnamefont {K.}~\bibnamefont {Watanabe}}, \emph
  {et~al.},\ }\href@noop {} {\bibfield  {journal} {\bibinfo  {journal}
  {Nature}\ }\textbf {\bibinfo {volume} {595}},\ \bibinfo {pages} {48}
  (\bibinfo {year} {2021}{\natexlab{a}})}\BibitemShut {NoStop}%
\bibitem [{\citenamefont {Smole{\'{n}}ski}\ \emph {et~al.}(2021)\citenamefont
  {Smole{\'{n}}ski}, \citenamefont {Dolgirev}, \citenamefont {Kuhlenkamp},
  \citenamefont {Popert}, \citenamefont {Shimazaki}, \citenamefont {Back},
  \citenamefont {Lu}, \citenamefont {Kroner}, \citenamefont {Watanabe},
  \citenamefont {Taniguchi}, \citenamefont {Esterlis}, \citenamefont {Demler},\
  and\ \citenamefont {Imamo{\u{g}}lu}}]{signature-wc-nature21}%
  \BibitemOpen
  \bibfield  {author} {\bibinfo {author} {\bibfnamefont {T.}~\bibnamefont
  {Smole{\'{n}}ski}}, \bibinfo {author} {\bibfnamefont {P.~E.}\ \bibnamefont
  {Dolgirev}}, \bibinfo {author} {\bibfnamefont {C.}~\bibnamefont
  {Kuhlenkamp}}, \bibinfo {author} {\bibfnamefont {A.}~\bibnamefont {Popert}},
  \bibinfo {author} {\bibfnamefont {Y.}~\bibnamefont {Shimazaki}}, \bibinfo
  {author} {\bibfnamefont {P.}~\bibnamefont {Back}}, \bibinfo {author}
  {\bibfnamefont {X.}~\bibnamefont {Lu}}, \bibinfo {author} {\bibfnamefont
  {M.}~\bibnamefont {Kroner}}, \bibinfo {author} {\bibfnamefont
  {K.}~\bibnamefont {Watanabe}}, \bibinfo {author} {\bibfnamefont
  {T.}~\bibnamefont {Taniguchi}}, \bibinfo {author} {\bibfnamefont
  {I.}~\bibnamefont {Esterlis}}, \bibinfo {author} {\bibfnamefont
  {E.}~\bibnamefont {Demler}},\ and\ \bibinfo {author} {\bibfnamefont
  {A.}~\bibnamefont {Imamo{\u{g}}lu}},\ }\href
  {https://doi.org/10.1038/s41586-021-03590-4} {\bibfield  {journal} {\bibinfo
  {journal} {Nature}\ }\textbf {\bibinfo {volume} {595}},\ \bibinfo {pages}
  {53} (\bibinfo {year} {2021})}\BibitemShut {NoStop}%
\bibitem [{\citenamefont {Tsui}\ \emph {et~al.}(2024)\citenamefont {Tsui},
  \citenamefont {He}, \citenamefont {Hu}, \citenamefont {Lake}, \citenamefont
  {Wang}, \citenamefont {Watanabe}, \citenamefont {Taniguchi}, \citenamefont
  {Zaletel},\ and\ \citenamefont {Yazdani}}]{yazdani-wc-nature24}%
  \BibitemOpen
  \bibfield  {author} {\bibinfo {author} {\bibfnamefont {Y.-C.}\ \bibnamefont
  {Tsui}}, \bibinfo {author} {\bibfnamefont {M.}~\bibnamefont {He}}, \bibinfo
  {author} {\bibfnamefont {Y.}~\bibnamefont {Hu}}, \bibinfo {author}
  {\bibfnamefont {E.}~\bibnamefont {Lake}}, \bibinfo {author} {\bibfnamefont
  {T.}~\bibnamefont {Wang}}, \bibinfo {author} {\bibfnamefont {K.}~\bibnamefont
  {Watanabe}}, \bibinfo {author} {\bibfnamefont {T.}~\bibnamefont {Taniguchi}},
  \bibinfo {author} {\bibfnamefont {M.~P.}\ \bibnamefont {Zaletel}},\ and\
  \bibinfo {author} {\bibfnamefont {A.}~\bibnamefont {Yazdani}},\ }\href@noop
  {} {\bibfield  {journal} {\bibinfo  {journal} {Nature}\ }\textbf {\bibinfo
  {volume} {628}},\ \bibinfo {pages} {287} (\bibinfo {year}
  {2024})}\BibitemShut {NoStop}%
\bibitem [{\citenamefont {Novoselov}\ \emph {et~al.}(2004)\citenamefont
  {Novoselov}, \citenamefont {Geim}, \citenamefont {Morozov}, \citenamefont
  {Jiang}, \citenamefont {Zhang}, \citenamefont {Dubonos}, \citenamefont
  {Grigorieva},\ and\ \citenamefont {Firsov}}]{graphene-science-04}%
  \BibitemOpen
  \bibfield  {author} {\bibinfo {author} {\bibfnamefont {K.~S.}\ \bibnamefont
  {Novoselov}}, \bibinfo {author} {\bibfnamefont {A.~K.}\ \bibnamefont {Geim}},
  \bibinfo {author} {\bibfnamefont {S.~V.}\ \bibnamefont {Morozov}}, \bibinfo
  {author} {\bibfnamefont {D.-e.}\ \bibnamefont {Jiang}}, \bibinfo {author}
  {\bibfnamefont {Y.}~\bibnamefont {Zhang}}, \bibinfo {author} {\bibfnamefont
  {S.~V.}\ \bibnamefont {Dubonos}}, \bibinfo {author} {\bibfnamefont {I.~V.}\
  \bibnamefont {Grigorieva}},\ and\ \bibinfo {author} {\bibfnamefont {A.~A.}\
  \bibnamefont {Firsov}},\ }\href@noop {} {\bibfield  {journal} {\bibinfo
  {journal} {Science}\ }\textbf {\bibinfo {volume} {306}},\ \bibinfo {pages}
  {666} (\bibinfo {year} {2004})}\BibitemShut {NoStop}%
\bibitem [{\citenamefont {Castro~Neto}\ \emph {et~al.}(2009)\citenamefont
  {Castro~Neto}, \citenamefont {Guinea}, \citenamefont {Peres}, \citenamefont
  {Novoselov},\ and\ \citenamefont {Geim}}]{graphene-rmp}%
  \BibitemOpen
  \bibfield  {author} {\bibinfo {author} {\bibfnamefont {A.~H.}\ \bibnamefont
  {Castro~Neto}}, \bibinfo {author} {\bibfnamefont {F.}~\bibnamefont {Guinea}},
  \bibinfo {author} {\bibfnamefont {N.~M.~R.}\ \bibnamefont {Peres}}, \bibinfo
  {author} {\bibfnamefont {K.~S.}\ \bibnamefont {Novoselov}},\ and\ \bibinfo
  {author} {\bibfnamefont {A.~K.}\ \bibnamefont {Geim}},\ }\href
  {https://doi.org/10.1103/RevModPhys.81.109} {\bibfield  {journal} {\bibinfo
  {journal} {Rev. Mod. Phys.}\ }\textbf {\bibinfo {volume} {81}},\ \bibinfo
  {pages} {109} (\bibinfo {year} {2009})}\BibitemShut {NoStop}%
\bibitem [{\citenamefont {Min}\ and\ \citenamefont
  {MacDonald}(2008)}]{min-multilayer-08}%
  \BibitemOpen
  \bibfield  {author} {\bibinfo {author} {\bibfnamefont {H.}~\bibnamefont
  {Min}}\ and\ \bibinfo {author} {\bibfnamefont {A.~H.}\ \bibnamefont
  {MacDonald}},\ }\href {https://doi.org/10.1143/PTPS.176.227} {\bibfield
  {journal} {\bibinfo  {journal} {Progress of Theoretical Physics Supplement}\
  }\textbf {\bibinfo {volume} {176}},\ \bibinfo {pages} {227} (\bibinfo {year}
  {2008})}\BibitemShut {NoStop}%
\bibitem [{\citenamefont {Dong}\ \emph
  {et~al.}(2024{\natexlab{a}})\citenamefont {Dong}, \citenamefont {Wang},
  \citenamefont {Wang}, \citenamefont {Soejima}, \citenamefont {Zaletel},
  \citenamefont {Vishwanath},\ and\ \citenamefont {Parker}}]{dong-AHC-prl24}%
  \BibitemOpen
  \bibfield  {author} {\bibinfo {author} {\bibfnamefont {J.}~\bibnamefont
  {Dong}}, \bibinfo {author} {\bibfnamefont {T.}~\bibnamefont {Wang}}, \bibinfo
  {author} {\bibfnamefont {T.}~\bibnamefont {Wang}}, \bibinfo {author}
  {\bibfnamefont {T.}~\bibnamefont {Soejima}}, \bibinfo {author} {\bibfnamefont
  {M.~P.}\ \bibnamefont {Zaletel}}, \bibinfo {author} {\bibfnamefont
  {A.}~\bibnamefont {Vishwanath}},\ and\ \bibinfo {author} {\bibfnamefont
  {D.~E.}\ \bibnamefont {Parker}},\ }\href
  {https://doi.org/10.1103/PhysRevLett.133.206503} {\bibfield  {journal}
  {\bibinfo  {journal} {Phys. Rev. Lett.}\ }\textbf {\bibinfo {volume} {133}},\
  \bibinfo {pages} {206503} (\bibinfo {year} {2024}{\natexlab{a}})}\BibitemShut
  {NoStop}%
\bibitem [{\citenamefont {Zhou}\ \emph {et~al.}(2024)\citenamefont {Zhou},
  \citenamefont {Yang},\ and\ \citenamefont {Zhang}}]{zhang-ahc-prl24}%
  \BibitemOpen
  \bibfield  {author} {\bibinfo {author} {\bibfnamefont {B.}~\bibnamefont
  {Zhou}}, \bibinfo {author} {\bibfnamefont {H.}~\bibnamefont {Yang}},\ and\
  \bibinfo {author} {\bibfnamefont {Y.-H.}\ \bibnamefont {Zhang}},\ }\href
  {https://doi.org/10.1103/PhysRevLett.133.206504} {\bibfield  {journal}
  {\bibinfo  {journal} {Phys. Rev. Lett.}\ }\textbf {\bibinfo {volume} {133}},\
  \bibinfo {pages} {206504} (\bibinfo {year} {2024})}\BibitemShut {NoStop}%
\bibitem [{\citenamefont {Tan}\ and\ \citenamefont
  {Devakul}(2024)}]{tan-AHC-prx24}%
  \BibitemOpen
  \bibfield  {author} {\bibinfo {author} {\bibfnamefont {T.}~\bibnamefont
  {Tan}}\ and\ \bibinfo {author} {\bibfnamefont {T.}~\bibnamefont {Devakul}},\
  }\href {https://doi.org/10.1103/PhysRevX.14.041040} {\bibfield  {journal}
  {\bibinfo  {journal} {Phys. Rev. X}\ }\textbf {\bibinfo {volume} {14}},\
  \bibinfo {pages} {041040} (\bibinfo {year} {2024})}\BibitemShut {NoStop}%
\bibitem [{\citenamefont {Dong}\ \emph
  {et~al.}(2024{\natexlab{b}})\citenamefont {Dong}, \citenamefont {Patri},\
  and\ \citenamefont {Senthil}}]{dong-AHC-prb24}%
  \BibitemOpen
  \bibfield  {author} {\bibinfo {author} {\bibfnamefont {Z.}~\bibnamefont
  {Dong}}, \bibinfo {author} {\bibfnamefont {A.~S.}\ \bibnamefont {Patri}},\
  and\ \bibinfo {author} {\bibfnamefont {T.}~\bibnamefont {Senthil}},\ }\href
  {https://doi.org/10.1103/PhysRevB.110.205130} {\bibfield  {journal} {\bibinfo
   {journal} {Phys. Rev. B}\ }\textbf {\bibinfo {volume} {110}},\ \bibinfo
  {pages} {205130} (\bibinfo {year} {2024}{\natexlab{b}})}\BibitemShut
  {NoStop}%
\bibitem [{\citenamefont {Rapisarda}\ and\ \citenamefont
  {Senatore}(1996)}]{rapisarda-dqmc-1996}%
  \BibitemOpen
  \bibfield  {author} {\bibinfo {author} {\bibfnamefont {F.}~\bibnamefont
  {Rapisarda}}\ and\ \bibinfo {author} {\bibfnamefont {G.}~\bibnamefont
  {Senatore}},\ }\href@noop {} {\bibfield  {journal} {\bibinfo  {journal}
  {Australian journal of physics}\ }\textbf {\bibinfo {volume} {49}},\ \bibinfo
  {pages} {161} (\bibinfo {year} {1996})}\BibitemShut {NoStop}%
\bibitem [{\citenamefont {Hedin}(1965)}]{hedin-gw-pr65}%
  \BibitemOpen
  \bibfield  {author} {\bibinfo {author} {\bibfnamefont {L.}~\bibnamefont
  {Hedin}},\ }\href {https://doi.org/10.1103/PhysRev.139.A796} {\bibfield
  {journal} {\bibinfo  {journal} {Phys. Rev.}\ }\textbf {\bibinfo {volume}
  {139}},\ \bibinfo {pages} {A796} (\bibinfo {year} {1965})}\BibitemShut
  {NoStop}%
\bibitem [{\citenamefont {Aryasetiawan}\ and\ \citenamefont
  {Gunnarsson}(1998)}]{aryasetiawan-gw-rpp98}%
  \BibitemOpen
  \bibfield  {author} {\bibinfo {author} {\bibfnamefont {F.}~\bibnamefont
  {Aryasetiawan}}\ and\ \bibinfo {author} {\bibfnamefont {O.}~\bibnamefont
  {Gunnarsson}},\ }\href {https://doi.org/10.1088/0034-4885/61/3/002}
  {\bibfield  {journal} {\bibinfo  {journal} {Reports on Progress in Physics}\
  }\textbf {\bibinfo {volume} {61}},\ \bibinfo {pages} {237} (\bibinfo {year}
  {1998})}\BibitemShut {NoStop}%
\bibitem [{\citenamefont {Reining}(2018)}]{reining-gw-wires18}%
  \BibitemOpen
  \bibfield  {author} {\bibinfo {author} {\bibfnamefont {L.}~\bibnamefont
  {Reining}},\ }\href {https://doi.org/https://doi.org/10.1002/wcms.1344}
  {\bibfield  {journal} {\bibinfo  {journal} {WIREs Computational Molecular
  Science}\ }\textbf {\bibinfo {volume} {8}},\ \bibinfo {pages} {e1344}
  (\bibinfo {year} {2018})}\BibitemShut {NoStop}%
\bibitem [{\citenamefont {Golze}\ \emph {et~al.}(2019)\citenamefont {Golze},
  \citenamefont {Dvorak},\ and\ \citenamefont {Rinke}}]{golze-gw-fc19}%
  \BibitemOpen
  \bibfield  {author} {\bibinfo {author} {\bibfnamefont {D.}~\bibnamefont
  {Golze}}, \bibinfo {author} {\bibfnamefont {M.}~\bibnamefont {Dvorak}},\ and\
  \bibinfo {author} {\bibfnamefont {P.}~\bibnamefont {Rinke}},\ }\href
  {https://doi.org/10.3389/fchem.2019.00377} {\bibfield  {journal} {\bibinfo
  {journal} {Frontiers in chemistry}\ }\textbf {\bibinfo {volume} {7}},\
  \bibinfo {pages} {377} (\bibinfo {year} {2019})}\BibitemShut {NoStop}%
\bibitem [{\citenamefont {Bohm}\ and\ \citenamefont
  {Pines}(1953)}]{bohm-rpa-pr53}%
  \BibitemOpen
  \bibfield  {author} {\bibinfo {author} {\bibfnamefont {D.}~\bibnamefont
  {Bohm}}\ and\ \bibinfo {author} {\bibfnamefont {D.}~\bibnamefont {Pines}},\
  }\href {https://doi.org/10.1103/PhysRev.92.609} {\bibfield  {journal}
  {\bibinfo  {journal} {Phys. Rev.}\ }\textbf {\bibinfo {volume} {92}},\
  \bibinfo {pages} {609} (\bibinfo {year} {1953})}\BibitemShut {NoStop}%
\bibitem [{\citenamefont {Gell-Mann}\ and\ \citenamefont
  {Brueckner}(1957)}]{gellmann-rpa-pr57}%
  \BibitemOpen
  \bibfield  {author} {\bibinfo {author} {\bibfnamefont {M.}~\bibnamefont
  {Gell-Mann}}\ and\ \bibinfo {author} {\bibfnamefont {K.~A.}\ \bibnamefont
  {Brueckner}},\ }\href {https://doi.org/10.1103/PhysRev.106.364} {\bibfield
  {journal} {\bibinfo  {journal} {Phys. Rev.}\ }\textbf {\bibinfo {volume}
  {106}},\ \bibinfo {pages} {364} (\bibinfo {year} {1957})}\BibitemShut
  {NoStop}%
\bibitem [{sup()}]{supp_info}%
  \BibitemOpen
  \href@noop {} {}\bibinfo {note} {See Supplemental Information (including
  Refs.~\cite{min-multilayer-08,jpliu-prx19,lu-nc23,hedin-gw-pr65,louie-ppa-prb86,aryasetiawan-gw-rpp98,rubio-rmp02,reining-gw-wires18,golze-gw-fc19,mpa-prb21,mpa-prb23,louie-ppa-prb86,hybertsen-ppa-prb89,needs-ppa-prl89,horsch-ppa-prb88,engel-ppa-prb93,macdonald-wc-prb91,fetter-book,bohm-rpa-pr53,gellmann-rpa-pr57,ren-rpa-2012,needs-wigner-qmc-prl09,rapisarda-dqmc-1996,graphene-arpes-science10,louie-gw-graphene-prl08,zhou-plasmaron-npjqm21})
  for: (a) detailed formalism for the Hartree-Fock calculations, (b) details of
  the $GW$ calculations, (c) detailed formalism multiple plasmon pole
  approximation, (d) details of the random phase approximation for correlation
  energy, (e) data fitting for critical Wigner-Seitz radius $r_s^*$ in 2DEG,
  and (f) more results about $n$-order Dirac fermion models for $n$=2, 3, 4, 5,
  and 6.}\BibitemShut {Stop}%
\bibitem [{\citenamefont {Fetter}\ and\ \citenamefont
  {Walecka}(2012)}]{fetter-book}%
  \BibitemOpen
  \bibfield  {author} {\bibinfo {author} {\bibfnamefont {A.~L.}\ \bibnamefont
  {Fetter}}\ and\ \bibinfo {author} {\bibfnamefont {J.~D.}\ \bibnamefont
  {Walecka}},\ }\href@noop {} {\emph {\bibinfo {title} {Quantum theory of
  many-particle systems}}}\ (\bibinfo  {publisher} {Courier Corporation},\
  \bibinfo {year} {2012})\BibitemShut {NoStop}%
\bibitem [{\citenamefont {Ren}\ \emph {et~al.}(2012)\citenamefont {Ren},
  \citenamefont {Rinke}, \citenamefont {Joas},\ and\ \citenamefont
  {Scheffler}}]{ren-rpa-2012}%
  \BibitemOpen
  \bibfield  {author} {\bibinfo {author} {\bibfnamefont {X.}~\bibnamefont
  {Ren}}, \bibinfo {author} {\bibfnamefont {P.}~\bibnamefont {Rinke}}, \bibinfo
  {author} {\bibfnamefont {C.}~\bibnamefont {Joas}},\ and\ \bibinfo {author}
  {\bibfnamefont {M.}~\bibnamefont {Scheffler}},\ }\href
  {https://doi.org/https://doi.org/10.1007/s10853-012-6570-4} {\bibfield
  {journal} {\bibinfo  {journal} {Journal of Materials Science}\ }\textbf
  {\bibinfo {volume} {47}},\ \bibinfo {pages} {7447} (\bibinfo {year}
  {2012})}\BibitemShut {NoStop}%
\bibitem [{\citenamefont {Hybertsen}\ and\ \citenamefont
  {Louie}(1986)}]{louie-ppa-prb86}%
  \BibitemOpen
  \bibfield  {author} {\bibinfo {author} {\bibfnamefont {M.~S.}\ \bibnamefont
  {Hybertsen}}\ and\ \bibinfo {author} {\bibfnamefont {S.~G.}\ \bibnamefont
  {Louie}},\ }\href {https://doi.org/10.1103/PhysRevB.34.5390} {\bibfield
  {journal} {\bibinfo  {journal} {Phys. Rev. B}\ }\textbf {\bibinfo {volume}
  {34}},\ \bibinfo {pages} {5390} (\bibinfo {year} {1986})}\BibitemShut
  {NoStop}%
\bibitem [{\citenamefont {Onida}\ \emph {et~al.}(2002)\citenamefont {Onida},
  \citenamefont {Reining},\ and\ \citenamefont {Rubio}}]{rubio-rmp02}%
  \BibitemOpen
  \bibfield  {author} {\bibinfo {author} {\bibfnamefont {G.}~\bibnamefont
  {Onida}}, \bibinfo {author} {\bibfnamefont {L.}~\bibnamefont {Reining}},\
  and\ \bibinfo {author} {\bibfnamefont {A.}~\bibnamefont {Rubio}},\ }\href
  {https://doi.org/10.1103/RevModPhys.74.601} {\bibfield  {journal} {\bibinfo
  {journal} {Rev. Mod. Phys.}\ }\textbf {\bibinfo {volume} {74}},\ \bibinfo
  {pages} {601} (\bibinfo {year} {2002})}\BibitemShut {NoStop}%
\bibitem [{\citenamefont {Leon}\ \emph {et~al.}(2021)\citenamefont {Leon},
  \citenamefont {Cardoso}, \citenamefont {Chiarotti}, \citenamefont {Varsano},
  \citenamefont {Molinari},\ and\ \citenamefont {Ferretti}}]{mpa-prb21}%
  \BibitemOpen
  \bibfield  {author} {\bibinfo {author} {\bibfnamefont {D.~A.}\ \bibnamefont
  {Leon}}, \bibinfo {author} {\bibfnamefont {C.}~\bibnamefont {Cardoso}},
  \bibinfo {author} {\bibfnamefont {T.}~\bibnamefont {Chiarotti}}, \bibinfo
  {author} {\bibfnamefont {D.}~\bibnamefont {Varsano}}, \bibinfo {author}
  {\bibfnamefont {E.}~\bibnamefont {Molinari}},\ and\ \bibinfo {author}
  {\bibfnamefont {A.}~\bibnamefont {Ferretti}},\ }\href
  {https://doi.org/10.1103/PhysRevB.104.115157} {\bibfield  {journal} {\bibinfo
   {journal} {Phys. Rev. B}\ }\textbf {\bibinfo {volume} {104}},\ \bibinfo
  {pages} {115157} (\bibinfo {year} {2021})}\BibitemShut {NoStop}%
\bibitem [{\citenamefont {Leon}\ \emph {et~al.}(2023)\citenamefont {Leon},
  \citenamefont {Ferretti}, \citenamefont {Varsano}, \citenamefont {Molinari},\
  and\ \citenamefont {Cardoso}}]{mpa-prb23}%
  \BibitemOpen
  \bibfield  {author} {\bibinfo {author} {\bibfnamefont {D.~A.}\ \bibnamefont
  {Leon}}, \bibinfo {author} {\bibfnamefont {A.}~\bibnamefont {Ferretti}},
  \bibinfo {author} {\bibfnamefont {D.}~\bibnamefont {Varsano}}, \bibinfo
  {author} {\bibfnamefont {E.}~\bibnamefont {Molinari}},\ and\ \bibinfo
  {author} {\bibfnamefont {C.}~\bibnamefont {Cardoso}},\ }\href
  {https://doi.org/10.1103/PhysRevB.107.155130} {\bibfield  {journal} {\bibinfo
   {journal} {Phys. Rev. B}\ }\textbf {\bibinfo {volume} {107}},\ \bibinfo
  {pages} {155130} (\bibinfo {year} {2023})}\BibitemShut {NoStop}%
\bibitem [{\citenamefont {Moon}\ and\ \citenamefont
  {Koshino}(2013)}]{moon-tbg-prb13}%
  \BibitemOpen
  \bibfield  {author} {\bibinfo {author} {\bibfnamefont {P.}~\bibnamefont
  {Moon}}\ and\ \bibinfo {author} {\bibfnamefont {M.}~\bibnamefont {Koshino}},\
  }\href@noop {} {\bibfield  {journal} {\bibinfo  {journal} {Physical Review
  B}\ }\textbf {\bibinfo {volume} {87}},\ \bibinfo {pages} {205404} (\bibinfo
  {year} {2013})}\BibitemShut {NoStop}%
\bibitem [{\citenamefont {McCann}\ and\ \citenamefont
  {Fal'ko}(2006)}]{blg-prl06}%
  \BibitemOpen
  \bibfield  {author} {\bibinfo {author} {\bibfnamefont {E.}~\bibnamefont
  {McCann}}\ and\ \bibinfo {author} {\bibfnamefont {V.~I.}\ \bibnamefont
  {Fal'ko}},\ }\href {https://doi.org/10.1103/PhysRevLett.96.086805} {\bibfield
   {journal} {\bibinfo  {journal} {Phys. Rev. Lett.}\ }\textbf {\bibinfo
  {volume} {96}},\ \bibinfo {pages} {086805} (\bibinfo {year}
  {2006})}\BibitemShut {NoStop}%
\bibitem [{\citenamefont {Bostwick}\ \emph {et~al.}(2010)\citenamefont
  {Bostwick}, \citenamefont {Speck}, \citenamefont {Seyller}, \citenamefont
  {Horn}, \citenamefont {Polini}, \citenamefont {Asgari}, \citenamefont
  {MacDonald},\ and\ \citenamefont {Rotenberg}}]{graphene-arpes-science10}%
  \BibitemOpen
  \bibfield  {author} {\bibinfo {author} {\bibfnamefont {A.}~\bibnamefont
  {Bostwick}}, \bibinfo {author} {\bibfnamefont {F.}~\bibnamefont {Speck}},
  \bibinfo {author} {\bibfnamefont {T.}~\bibnamefont {Seyller}}, \bibinfo
  {author} {\bibfnamefont {K.}~\bibnamefont {Horn}}, \bibinfo {author}
  {\bibfnamefont {M.}~\bibnamefont {Polini}}, \bibinfo {author} {\bibfnamefont
  {R.}~\bibnamefont {Asgari}}, \bibinfo {author} {\bibfnamefont {A.~H.}\
  \bibnamefont {MacDonald}},\ and\ \bibinfo {author} {\bibfnamefont
  {E.}~\bibnamefont {Rotenberg}},\ }\href
  {https://doi.org/10.1126/science.1186489} {\bibfield  {journal} {\bibinfo
  {journal} {Science}\ }\textbf {\bibinfo {volume} {328}},\ \bibinfo {pages}
  {999} (\bibinfo {year} {2010})}\BibitemShut {NoStop}%
\bibitem [{\citenamefont {Trevisanutto}\ \emph {et~al.}(2008)\citenamefont
  {Trevisanutto}, \citenamefont {Giorgetti}, \citenamefont {Reining},
  \citenamefont {Ladisa},\ and\ \citenamefont
  {Olevano}}]{louie-gw-graphene-prl08}%
  \BibitemOpen
  \bibfield  {author} {\bibinfo {author} {\bibfnamefont {P.~E.}\ \bibnamefont
  {Trevisanutto}}, \bibinfo {author} {\bibfnamefont {C.}~\bibnamefont
  {Giorgetti}}, \bibinfo {author} {\bibfnamefont {L.}~\bibnamefont {Reining}},
  \bibinfo {author} {\bibfnamefont {M.}~\bibnamefont {Ladisa}},\ and\ \bibinfo
  {author} {\bibfnamefont {V.}~\bibnamefont {Olevano}},\ }\href
  {https://doi.org/10.1103/PhysRevLett.101.226405} {\bibfield  {journal}
  {\bibinfo  {journal} {Phys. Rev. Lett.}\ }\textbf {\bibinfo {volume} {101}},\
  \bibinfo {pages} {226405} (\bibinfo {year} {2008})}\BibitemShut {NoStop}%
\bibitem [{\citenamefont {Zhang}\ \emph {et~al.}(2021)\citenamefont {Zhang},
  \citenamefont {Wang}, \citenamefont {Wang}, \citenamefont {Lu}, \citenamefont
  {Li}, \citenamefont {Bao}, \citenamefont {Deng}, \citenamefont {Zhang},
  \citenamefont {Yao}, \citenamefont {Chen}, \citenamefont {Fedorov},
  \citenamefont {Denlinger}, \citenamefont {Watanabe}, \citenamefont
  {Taniguchi}, \citenamefont {Zhang},\ and\ \citenamefont
  {Zhou}}]{zhou-plasmaron-npjqm21}%
  \BibitemOpen
  \bibfield  {author} {\bibinfo {author} {\bibfnamefont {H.}~\bibnamefont
  {Zhang}}, \bibinfo {author} {\bibfnamefont {S.}~\bibnamefont {Wang}},
  \bibinfo {author} {\bibfnamefont {E.}~\bibnamefont {Wang}}, \bibinfo {author}
  {\bibfnamefont {X.}~\bibnamefont {Lu}}, \bibinfo {author} {\bibfnamefont
  {Q.}~\bibnamefont {Li}}, \bibinfo {author} {\bibfnamefont {C.}~\bibnamefont
  {Bao}}, \bibinfo {author} {\bibfnamefont {K.}~\bibnamefont {Deng}}, \bibinfo
  {author} {\bibfnamefont {H.}~\bibnamefont {Zhang}}, \bibinfo {author}
  {\bibfnamefont {W.}~\bibnamefont {Yao}}, \bibinfo {author} {\bibfnamefont
  {G.}~\bibnamefont {Chen}}, \bibinfo {author} {\bibfnamefont {A.~V.}\
  \bibnamefont {Fedorov}}, \bibinfo {author} {\bibfnamefont {J.~D.}\
  \bibnamefont {Denlinger}}, \bibinfo {author} {\bibfnamefont {K.}~\bibnamefont
  {Watanabe}}, \bibinfo {author} {\bibfnamefont {T.}~\bibnamefont {Taniguchi}},
  \bibinfo {author} {\bibfnamefont {G.}~\bibnamefont {Zhang}},\ and\ \bibinfo
  {author} {\bibfnamefont {S.}~\bibnamefont {Zhou}},\ }\href
  {https://doi.org/10.1038/s41535-021-00386-7} {\bibfield  {journal} {\bibinfo
  {journal} {npj Quantum Materials}\ }\textbf {\bibinfo {volume} {6}},\
  \bibinfo {pages} {83} (\bibinfo {year} {2021})}\BibitemShut {NoStop}%
\bibitem [{\citenamefont {Han}\ \emph {et~al.}(2024)\citenamefont {Han},
  \citenamefont {Lu}, \citenamefont {Scuri}, \citenamefont {Sung},
  \citenamefont {Wang}, \citenamefont {Han}, \citenamefont {Watanabe},
  \citenamefont {Taniguchi}, \citenamefont {Park},\ and\ \citenamefont
  {Ju}}]{ju-chern-natnano2023}%
  \BibitemOpen
  \bibfield  {author} {\bibinfo {author} {\bibfnamefont {T.}~\bibnamefont
  {Han}}, \bibinfo {author} {\bibfnamefont {Z.}~\bibnamefont {Lu}}, \bibinfo
  {author} {\bibfnamefont {G.}~\bibnamefont {Scuri}}, \bibinfo {author}
  {\bibfnamefont {J.}~\bibnamefont {Sung}}, \bibinfo {author} {\bibfnamefont
  {J.}~\bibnamefont {Wang}}, \bibinfo {author} {\bibfnamefont {T.}~\bibnamefont
  {Han}}, \bibinfo {author} {\bibfnamefont {K.}~\bibnamefont {Watanabe}},
  \bibinfo {author} {\bibfnamefont {T.}~\bibnamefont {Taniguchi}}, \bibinfo
  {author} {\bibfnamefont {H.}~\bibnamefont {Park}},\ and\ \bibinfo {author}
  {\bibfnamefont {L.}~\bibnamefont {Ju}},\ }\href
  {https://doi.org/10.1038/s41565-023-01520-1} {\bibfield  {journal} {\bibinfo
  {journal} {Nature Nanotechnology}\ }\textbf {\bibinfo {volume} {19}},\
  \bibinfo {pages} {181} (\bibinfo {year} {2024})}\BibitemShut {NoStop}%
\bibitem [{\citenamefont {Zhou}\ \emph
  {et~al.}(2021{\natexlab{b}})\citenamefont {Zhou}, \citenamefont {Xie},
  \citenamefont {Ghazaryan}, \citenamefont {Holder}, \citenamefont {Ehrets},
  \citenamefont {Spanton}, \citenamefont {Taniguchi}, \citenamefont {Watanabe},
  \citenamefont {Berg}, \citenamefont {Serbyn} \emph
  {et~al.}}]{zhou-trilayer-nature21}%
  \BibitemOpen
  \bibfield  {author} {\bibinfo {author} {\bibfnamefont {H.}~\bibnamefont
  {Zhou}}, \bibinfo {author} {\bibfnamefont {T.}~\bibnamefont {Xie}}, \bibinfo
  {author} {\bibfnamefont {A.}~\bibnamefont {Ghazaryan}}, \bibinfo {author}
  {\bibfnamefont {T.}~\bibnamefont {Holder}}, \bibinfo {author} {\bibfnamefont
  {J.~R.}\ \bibnamefont {Ehrets}}, \bibinfo {author} {\bibfnamefont {E.~M.}\
  \bibnamefont {Spanton}}, \bibinfo {author} {\bibfnamefont {T.}~\bibnamefont
  {Taniguchi}}, \bibinfo {author} {\bibfnamefont {K.}~\bibnamefont {Watanabe}},
  \bibinfo {author} {\bibfnamefont {E.}~\bibnamefont {Berg}}, \bibinfo {author}
  {\bibfnamefont {M.}~\bibnamefont {Serbyn}}, \emph {et~al.},\ }\href@noop {}
  {\bibfield  {journal} {\bibinfo  {journal} {Nature}\ }\textbf {\bibinfo
  {volume} {598}},\ \bibinfo {pages} {429} (\bibinfo {year}
  {2021}{\natexlab{b}})}\BibitemShut {NoStop}%
\bibitem [{\citenamefont {Zeng}\ \emph {et~al.}(2024)\citenamefont {Zeng},
  \citenamefont {Guerci}, \citenamefont {Cr\'epel}, \citenamefont {Millis},\
  and\ \citenamefont {Cano}}]{cano-ahc-prl24}%
  \BibitemOpen
  \bibfield  {author} {\bibinfo {author} {\bibfnamefont {Y.}~\bibnamefont
  {Zeng}}, \bibinfo {author} {\bibfnamefont {D.}~\bibnamefont {Guerci}},
  \bibinfo {author} {\bibfnamefont {V.}~\bibnamefont {Cr\'epel}}, \bibinfo
  {author} {\bibfnamefont {A.~J.}\ \bibnamefont {Millis}},\ and\ \bibinfo
  {author} {\bibfnamefont {J.}~\bibnamefont {Cano}},\ }\href
  {https://doi.org/10.1103/PhysRevLett.132.236601} {\bibfield  {journal}
  {\bibinfo  {journal} {Phys. Rev. Lett.}\ }\textbf {\bibinfo {volume} {132}},\
  \bibinfo {pages} {236601} (\bibinfo {year} {2024})}\BibitemShut {NoStop}%
\bibitem [{\citenamefont {Li}\ \emph {et~al.}(2024)\citenamefont {Li},
  \citenamefont {Xu}, \citenamefont {Li}, \citenamefont {Li}, \citenamefont
  {Li}, \citenamefont {Watanabe}, \citenamefont {Taniguchi}, \citenamefont
  {Tong}, \citenamefont {Shen}, \citenamefont {Lu}, \citenamefont {Jia},
  \citenamefont {Wu}, \citenamefont {Liu},\ and\ \citenamefont
  {Li}}]{li-blg-nature24}%
  \BibitemOpen
  \bibfield  {author} {\bibinfo {author} {\bibfnamefont {C.}~\bibnamefont
  {Li}}, \bibinfo {author} {\bibfnamefont {F.}~\bibnamefont {Xu}}, \bibinfo
  {author} {\bibfnamefont {B.}~\bibnamefont {Li}}, \bibinfo {author}
  {\bibfnamefont {J.}~\bibnamefont {Li}}, \bibinfo {author} {\bibfnamefont
  {G.}~\bibnamefont {Li}}, \bibinfo {author} {\bibfnamefont {K.}~\bibnamefont
  {Watanabe}}, \bibinfo {author} {\bibfnamefont {T.}~\bibnamefont {Taniguchi}},
  \bibinfo {author} {\bibfnamefont {B.}~\bibnamefont {Tong}}, \bibinfo {author}
  {\bibfnamefont {J.}~\bibnamefont {Shen}}, \bibinfo {author} {\bibfnamefont
  {L.}~\bibnamefont {Lu}}, \bibinfo {author} {\bibfnamefont {J.}~\bibnamefont
  {Jia}}, \bibinfo {author} {\bibfnamefont {F.}~\bibnamefont {Wu}}, \bibinfo
  {author} {\bibfnamefont {X.}~\bibnamefont {Liu}},\ and\ \bibinfo {author}
  {\bibfnamefont {T.}~\bibnamefont {Li}},\ }\href
  {https://doi.org/10.1038/s41586-024-07584-w} {\bibfield  {journal} {\bibinfo
  {journal} {Nature}\ }\textbf {\bibinfo {volume} {631}},\ \bibinfo {pages}
  {300} (\bibinfo {year} {2024})}\BibitemShut {NoStop}%
\bibitem [{\citenamefont {Zhou}\ \emph {et~al.}(2022)\citenamefont {Zhou},
  \citenamefont {Holleis}, \citenamefont {Saito}, \citenamefont {Cohen},
  \citenamefont {Huynh}, \citenamefont {Patterson}, \citenamefont {Yang},
  \citenamefont {Taniguchi}, \citenamefont {Watanabe},\ and\ \citenamefont
  {Young}}]{young-blg-science22}%
  \BibitemOpen
  \bibfield  {author} {\bibinfo {author} {\bibfnamefont {H.}~\bibnamefont
  {Zhou}}, \bibinfo {author} {\bibfnamefont {L.}~\bibnamefont {Holleis}},
  \bibinfo {author} {\bibfnamefont {Y.}~\bibnamefont {Saito}}, \bibinfo
  {author} {\bibfnamefont {L.}~\bibnamefont {Cohen}}, \bibinfo {author}
  {\bibfnamefont {W.}~\bibnamefont {Huynh}}, \bibinfo {author} {\bibfnamefont
  {C.~L.}\ \bibnamefont {Patterson}}, \bibinfo {author} {\bibfnamefont
  {F.}~\bibnamefont {Yang}}, \bibinfo {author} {\bibfnamefont {T.}~\bibnamefont
  {Taniguchi}}, \bibinfo {author} {\bibfnamefont {K.}~\bibnamefont
  {Watanabe}},\ and\ \bibinfo {author} {\bibfnamefont {A.~F.}\ \bibnamefont
  {Young}},\ }\href {https://doi.org/10.1126/science.abm8386} {\bibfield
  {journal} {\bibinfo  {journal} {Science}\ }\textbf {\bibinfo {volume}
  {375}},\ \bibinfo {pages} {774} (\bibinfo {year} {2022})},\ \Eprint
  {https://arxiv.org/abs/https://www.science.org/doi/pdf/10.1126/science.abm8386}
  {https://www.science.org/doi/pdf/10.1126/science.abm8386} \BibitemShut
  {NoStop}%
\bibitem [{\citenamefont {Seiler}\ \emph {et~al.}(2022)\citenamefont {Seiler},
  \citenamefont {Geisenhof}, \citenamefont {Winterer}, \citenamefont
  {Watanabe}, \citenamefont {Taniguchi}, \citenamefont {Xu}, \citenamefont
  {Zhang},\ and\ \citenamefont {Weitz}}]{weitz-blg-nature22}%
  \BibitemOpen
  \bibfield  {author} {\bibinfo {author} {\bibfnamefont {A.~M.}\ \bibnamefont
  {Seiler}}, \bibinfo {author} {\bibfnamefont {F.~R.}\ \bibnamefont
  {Geisenhof}}, \bibinfo {author} {\bibfnamefont {F.}~\bibnamefont {Winterer}},
  \bibinfo {author} {\bibfnamefont {K.}~\bibnamefont {Watanabe}}, \bibinfo
  {author} {\bibfnamefont {T.}~\bibnamefont {Taniguchi}}, \bibinfo {author}
  {\bibfnamefont {T.}~\bibnamefont {Xu}}, \bibinfo {author} {\bibfnamefont
  {F.}~\bibnamefont {Zhang}},\ and\ \bibinfo {author} {\bibfnamefont {R.~T.}\
  \bibnamefont {Weitz}},\ }\href {https://doi.org/10.1038/s41586-022-04937-1}
  {\bibfield  {journal} {\bibinfo  {journal} {Nature}\ }\textbf {\bibinfo
  {volume} {608}},\ \bibinfo {pages} {298} (\bibinfo {year}
  {2022})}\BibitemShut {NoStop}%
\bibitem [{\citenamefont {Liu}\ \emph {et~al.}(2019)\citenamefont {Liu},
  \citenamefont {Ma}, \citenamefont {Gao},\ and\ \citenamefont
  {Dai}}]{jpliu-prx19}%
  \BibitemOpen
  \bibfield  {author} {\bibinfo {author} {\bibfnamefont {J.}~\bibnamefont
  {Liu}}, \bibinfo {author} {\bibfnamefont {Z.}~\bibnamefont {Ma}}, \bibinfo
  {author} {\bibfnamefont {J.}~\bibnamefont {Gao}},\ and\ \bibinfo {author}
  {\bibfnamefont {X.}~\bibnamefont {Dai}},\ }\href
  {https://doi.org/10.1103/PhysRevX.9.031021} {\bibfield  {journal} {\bibinfo
  {journal} {Phys. Rev. X}\ }\textbf {\bibinfo {volume} {9}},\ \bibinfo {pages}
  {031021} (\bibinfo {year} {2019})}\BibitemShut {NoStop}%
\bibitem [{\citenamefont {Lu}\ \emph {et~al.}(2023)\citenamefont {Lu},
  \citenamefont {Zhang}, \citenamefont {Wang}, \citenamefont {Gao},
  \citenamefont {Yang}, \citenamefont {Guo}, \citenamefont {Gao}, \citenamefont
  {Ye}, \citenamefont {Han},\ and\ \citenamefont {Liu}}]{lu-nc23}%
  \BibitemOpen
  \bibfield  {author} {\bibinfo {author} {\bibfnamefont {X.}~\bibnamefont
  {Lu}}, \bibinfo {author} {\bibfnamefont {S.}~\bibnamefont {Zhang}}, \bibinfo
  {author} {\bibfnamefont {Y.}~\bibnamefont {Wang}}, \bibinfo {author}
  {\bibfnamefont {X.}~\bibnamefont {Gao}}, \bibinfo {author} {\bibfnamefont
  {K.}~\bibnamefont {Yang}}, \bibinfo {author} {\bibfnamefont {Z.}~\bibnamefont
  {Guo}}, \bibinfo {author} {\bibfnamefont {Y.}~\bibnamefont {Gao}}, \bibinfo
  {author} {\bibfnamefont {Y.}~\bibnamefont {Ye}}, \bibinfo {author}
  {\bibfnamefont {Z.}~\bibnamefont {Han}},\ and\ \bibinfo {author}
  {\bibfnamefont {J.}~\bibnamefont {Liu}},\ }\href
  {https://doi.org/10.1038/s41467-023-41293-8} {\bibfield  {journal} {\bibinfo
  {journal} {Nature Communications}\ }\textbf {\bibinfo {volume} {14}},\
  \bibinfo {pages} {5550} (\bibinfo {year} {2023})}\BibitemShut {NoStop}%
\bibitem [{\citenamefont {Zhang}\ \emph {et~al.}(1989)\citenamefont {Zhang},
  \citenamefont {Tom\'anek}, \citenamefont {Cohen}, \citenamefont {Louie},\
  and\ \citenamefont {Hybertsen}}]{hybertsen-ppa-prb89}%
  \BibitemOpen
  \bibfield  {author} {\bibinfo {author} {\bibfnamefont {S.~B.}\ \bibnamefont
  {Zhang}}, \bibinfo {author} {\bibfnamefont {D.}~\bibnamefont {Tom\'anek}},
  \bibinfo {author} {\bibfnamefont {M.~L.}\ \bibnamefont {Cohen}}, \bibinfo
  {author} {\bibfnamefont {S.~G.}\ \bibnamefont {Louie}},\ and\ \bibinfo
  {author} {\bibfnamefont {M.~S.}\ \bibnamefont {Hybertsen}},\ }\href
  {https://doi.org/10.1103/PhysRevB.40.3162} {\bibfield  {journal} {\bibinfo
  {journal} {Phys. Rev. B}\ }\textbf {\bibinfo {volume} {40}},\ \bibinfo
  {pages} {3162} (\bibinfo {year} {1989})}\BibitemShut {NoStop}%
\bibitem [{\citenamefont {Godby}\ and\ \citenamefont
  {Needs}(1989)}]{needs-ppa-prl89}%
  \BibitemOpen
  \bibfield  {author} {\bibinfo {author} {\bibfnamefont {R.~W.}\ \bibnamefont
  {Godby}}\ and\ \bibinfo {author} {\bibfnamefont {R.~J.}\ \bibnamefont
  {Needs}},\ }\href {https://doi.org/10.1103/PhysRevLett.62.1169} {\bibfield
  {journal} {\bibinfo  {journal} {Phys. Rev. Lett.}\ }\textbf {\bibinfo
  {volume} {62}},\ \bibinfo {pages} {1169} (\bibinfo {year}
  {1989})}\BibitemShut {NoStop}%
\bibitem [{\citenamefont {von~der Linden}\ and\ \citenamefont
  {Horsch}(1988)}]{horsch-ppa-prb88}%
  \BibitemOpen
  \bibfield  {author} {\bibinfo {author} {\bibfnamefont {W.}~\bibnamefont
  {von~der Linden}}\ and\ \bibinfo {author} {\bibfnamefont {P.}~\bibnamefont
  {Horsch}},\ }\href {https://doi.org/10.1103/PhysRevB.37.8351} {\bibfield
  {journal} {\bibinfo  {journal} {Phys. Rev. B}\ }\textbf {\bibinfo {volume}
  {37}},\ \bibinfo {pages} {8351} (\bibinfo {year} {1988})}\BibitemShut
  {NoStop}%
\bibitem [{\citenamefont {Engel}\ and\ \citenamefont
  {Farid}(1993)}]{engel-ppa-prb93}%
  \BibitemOpen
  \bibfield  {author} {\bibinfo {author} {\bibfnamefont {G.~E.}\ \bibnamefont
  {Engel}}\ and\ \bibinfo {author} {\bibfnamefont {B.}~\bibnamefont {Farid}},\
  }\href {https://doi.org/10.1103/PhysRevB.47.15931} {\bibfield  {journal}
  {\bibinfo  {journal} {Phys. Rev. B}\ }\textbf {\bibinfo {volume} {47}},\
  \bibinfo {pages} {15931} (\bibinfo {year} {1993})}\BibitemShut {NoStop}%
\bibitem [{\citenamefont {C\^ot\'e}\ and\ \citenamefont
  {MacDonald}(1991)}]{macdonald-wc-prb91}%
  \BibitemOpen
  \bibfield  {author} {\bibinfo {author} {\bibfnamefont {R.}~\bibnamefont
  {C\^ot\'e}}\ and\ \bibinfo {author} {\bibfnamefont {A.~H.}\ \bibnamefont
  {MacDonald}},\ }\href {https://doi.org/10.1103/PhysRevB.44.8759} {\bibfield
  {journal} {\bibinfo  {journal} {Phys. Rev. B}\ }\textbf {\bibinfo {volume}
  {44}},\ \bibinfo {pages} {8759} (\bibinfo {year} {1991})}\BibitemShut
  {NoStop}%
\end{thebibliography}%

\clearpage
\begin{widetext}
\begin{center}
	\textbf{\large Supplemental Material for ``Correlation stabilized anomalous Hall crystal in bilayer graphene"}
\end{center}

\maketitle
\setcounter{equation}{0}
\setcounter{figure}{0}
\setcounter{table}{0}
\setcounter{page}{1}
\makeatletter

\section{Hartree-Fock approximation for two-dimensional systems}

The Hartree-Fock (HF) approximation is a standard method to treat interacting electronic systems. Our HF framework is implemented in the plane-wave basis, where an electron annihilation operator $\cop_{\lambda}(\bk)$ with  flavor index $\lambda$ and wavevector $\bk$ (expanded around some low-energy valley),  can be re-written as:
\begin{equation}
	\cop_{\lambda,\bG}(\btk)\equiv\cop_{\lambda}(\bk),
\end{equation}
where $\lambda \equiv (\mu, \alpha, \sigma)$ is a composite flavor index denoting the (possible) valley, (possible) sublattice, and spin degrees of freedom. Wavevector $\bk$ is expanded around some low-energy valley, and is further decomposed into a reciprocal vector $\bG$  and  a wavevector $\btk$ within the Brillouin zone of the corresponding lattice. 
The non-interacting Hamiltonian  of the system can be generally expressed as:
\begin{equation}
	H^0=\sum_{\lambda,\lambda',\btk,\bG}H^0_{\lambda,\lambda'}(\btk+\bG)\,\cop^\dagger_{\lambda,\bG}(\btk)\cop_{\lambda',\bG}(\btk).
\end{equation}
For the conventional two-dimensional electron gas (2DEG) system, the kinetic energy is simply given by:
\begin{equation}   
	H_{\rm{2DEG}}^0(\bk) = \frac{\hbar^2 \bk^2}{2m^*}\,,
\end{equation}
where $m^*$ is the effective mass of the electrons, $\bk=\btk+\bG$. For $n$-order Dirac fermion models, which describe $n$-layer graphene system, the kinetic energy term is \cite{min-multilayer-08,jpliu-prx19}:
\begin{equation}
	H^0_{\mu,n}(\bk) = -
	\begin{pmatrix}
		\Delta & t_\perp(\nu_\mu^\dagger)^n \\ 
		t_\perp(\nu_\mu)^n & -\Delta
	\end{pmatrix},
\end{equation}
where $\nu_\mu = \hbar v_F (\mu k_x + ik_y)/t_\perp$, $v_F$ is the Fermi velocity, and $t_\perp$ is the interlayer hopping amplitude, $\mu=\mp$ denotes $K/K'$ valley of graphene. $\Delta$ denotes the Dirac-fermion mass  which physically originates from vertical electric field applied to rhombohedral multilayer graphene. Unless specified otherwise, we set $\hbar v_F=5.25\,\text{eV}\cdot\text{\AA}$, $t_{\perp}=0.34\,$eV, and $\Delta=0.1\,$eV throughout the paper.

In this work, we consider the Coulomb interaction only in the intravalley form, neglecting the intervalley interactions, which is usually exponentially smaller than the intravalley one. This assumption simplifies the interaction Hamiltonian and is justified because intravalley interactions are generally the dominant contributions at low energies. The interaction Hamiltonian within this approximation is written as \cite{lu-nc23}:
\begin{equation}
	H_{C}^{\text{intra}} = \frac{1}{2N_s} \sum_{\lambda, \lambda'} \sum_{\btk, \btk', \btq} \sum_{\bG, \bG', \bQ} V(\btq + \bQ) 
	\cop^{\dagger}_{\lambda, \bG+\bQ}(\btk + \btq) 
	\cop^{\dagger}_{\lambda', \bG'-\bQ}(\btk' - \btq) 
	\cop_{\lambda', \bG'}(\btk') 
	\cop_{\lambda, \bG}(\btk).
\end{equation}
Here, $N_s$ denotes the total number of unit cells (of the presumable Wigner crystal state) in the system. Assuming that the ground state always preserves lattice translational symmetry, the expectation value of the density operator is expressed as:
\begin{align}
	\langle \cop^\dagger_{\lambda,\bG+\bQ}(\btk+\btq)\cop_{\lambda',\bG}(\btk') \rangle
	=\langle \cop^\dagger_{\mu,\alpha,\bG+\bQ}(\btk')\cop_{\mu',\alpha',\bG}(\btk') \rangle\delta_{\btk+\btq,\btk'}\delta_{\sigma,\sigma'}.
\end{align}
Then, the Hartree term reads:
\begin{equation}
	V^{\text{H}} = \frac{1}{N_s} \sum_{\btk, \btk'} \sum_{\lambda, \lambda'} \sum_{\bG, \bG', \bQ} V(\bQ) 
	\langle 
	\cop_{\lambda', \bG'-\bQ}^{\dagger}(\btk') 
	\cop_{\lambda', \bG'}(\btk') \rangle 
	\cop_{\lambda, \bG+\bQ}^{\dagger}(\btk) 
	\cop_{\lambda, \bG}(\btk),
\end{equation}
and the Fock term reads:
\begin{equation}
	\begin{aligned}
		V_{\text{F}} &= -\frac{1}{N_s} \sum_{\btk, \btk'} \sum_{\lambda, \lambda'} \sum_{\bG, \bG', \bQ} V(\btk' - \btk + \bQ) 
		\langle 
		\cop_{\lambda, \bG+\bQ}^{\dagger}(\btk') 
		\cop_{\lambda', \bG'}(\btk') \rangle 
		\cop_{\lambda', \bG'-\bQ}^{\dagger}(\btk) 
		\cop_{\lambda, \bG}(\btk) \\
		&= -\frac{1}{N_s} \sum_{\btk, \btk'} \sum_{\lambda, \lambda'} \sum_{\bG, \bG', \bQ} V(\btk' - \btk+ \bG' - \bG + \bQ ) 
		\langle 
		\cop_{\lambda, \bG'+\bQ}^{\dagger}(\btk') 
		\cop_{\lambda', \bG'}(\btk') \rangle 
		\cop_{\lambda', \bG-\bQ}^{\dagger}(\btk) 
		\cop_{\lambda, \bG}(\btk).
	\end{aligned}
\end{equation}

In these equations, $\langle \dots \rangle$ represents the expectation value of the operator in the many-body ground state $\ket{\Psi}^0$, which is assumed to be a Slater-determinant state in the HF framework. To achieve a self-consistent solution, the HF calculations are carried out iteratively for different superlattice constants $L_s$. The background dielectric constant is set to $\epsilon_r = 4$. To initialize the HF self-consistent loop, the expectation values of density operators $\langle \cop_{\lambda, \bG'-\bQ}^{\dagger}(\btk') \cop_{\lambda, \bG'}(\btk') \rangle$ are chosen to reflect either a spontaneous charge order with non-zero Fourier component at $\bQ\neq\mathbf{0}$, corresponding to the Wigner crystal (WC) state; or it is  set to zero, corresponding to the Fermi liquid (FL) state. We compare the total energy $E_{\text{total}}=E_{\text{kinetic}}+E_{\text{HF}}$ of these two types of states to determine the genuine ground state. Here $E_{\text{kinetic}}$ is the kinetic energy, and $E_{\text{HF}}$ is the HF energy including both Hartree and exchange energies. In our calculations, including HF and $GW$+RPA, a $9 \times 9$ mesh of reciprocal lattice points (centered at the $\Gamma$ point) has been used for the 2DEG system, while a $5 \times 5$ mesh is used for $n$-order Dirac fermion systems. The mini Brillouin zone (of the presumable WC state) is sampled by an $18 \times 18$ $\bk$-mesh. 

\section{$GW$ approximation for two-dimensional systems}

The $GW$ approximation provides a framework for incorporating electron-electron interactions beyond the mean-field HF theory \cite{hedin-gw-pr65,louie-ppa-prb86,aryasetiawan-gw-rpp98,rubio-rmp02,reining-gw-wires18,golze-gw-fc19}. It improves the description of quasiparticle (QP) energies and the single-particle energy spectrum by including the effects of dynamic screening. The key equations in the $GW$ formalism are known as Hedin's equations, which describe a set of self-consistent equations of self-energy $\Sigma$, Green's function $G$,  vertex function $\Gamma$,  polarization propagator $P$, and  screened Coulomb interaction $W$. For simplicity, here we use Arabic number as a short-hand notation for spatial ($\br$) and temporal ($t$) coordinate, e.g., $1\equiv(\br_1, t_1)$. Then, with such short-hand notations, Hedin's equations are given by:
\begin{equation}
	\Sigma(12) = i \int \text{d}3 \, G(13) W(14) \Gamma(342),
\end{equation}
\begin{equation}
	G(12) = G_0(12) + \int \text{d}3 \, G_0(13) \Sigma(34) G(42),
\end{equation}
\begin{equation}
	\Gamma(123) = \delta(12) \delta(13) + \int \text{d}4 \, \text{d}5 \, \frac{\delta \Sigma(12)}{\delta G(45)} G(46) G(75) \Gamma(673),
\end{equation}
\begin{equation}
	P(12) = -i \int \text{d}3 \, \text{d}4 \, G(13) G(42) \Gamma(342),
\end{equation}
\begin{equation}
	W(12) = V(12) + \int \text{d}3 \, V(13) P(34) W(42).
\end{equation}

These equations form the basis of many-body perturbation theory, linking the Green's function, self-energy, and screened interaction in a self-consistent framework. To apply the $GW$ approximation, we assume a simple form of vertex function, $\Gamma(123) = \delta(12) \delta(13)$, which leads to the "bare vertex" approximation:
\begin{equation}
	\Gamma(123) = \delta(12) \delta(13).
\end{equation}
Under this approximation, the self-energy simplifies to:
\begin{equation}
	\Sigma(12) = i G(12) W(12).
\end{equation}
Similarly, the polarization reduces to:
\begin{equation}
	P(12) = -i G(12) G(21).
\end{equation}
To describe the screened Coulomb interaction $W$ in terms of the dielectric function $\epsilon$, we express $V$ as:
\begin{equation}
	V(12)=\int\text{d}{3}\epsilon(13)W(32).
\end{equation}
The dielectric function $\epsilon$ can be formulated as:
\begin{equation}
	\epsilon(12)=\delta(12)-\int\text{d}{3}V(13)P(32).
\end{equation}

We continue to perform a Fourier transform from time domain to frequency domain to handle the time-dependent components. Now we go back to the usual notations where $\br$ ($\br'$) denote real-space coordinate, and $\omega$ denote frequency. The non-interacting Green's function $G_0$ is given by:
\begin{align}
	G_0(\br,\br',\omega)=\sum_{n\btk}\frac{\psi_{n\btk}(\br)\psi^*_{n\btk}(\br')}{\omega-\varepsilon_{n\btk}+i\delta\mathrm{sgn}(\varepsilon_{n\btk}-\varepsilon_F)},
\end{align}
where $\psi_{n\btk}$ are the HF single-particle wave functions, $\varepsilon_{n\btk}$ are the HF eigenvalues, and $\varepsilon_F$ is the Fermi energy.
The non-interacting charge polarizability $\chi^0$, which characterizes the linear response of the system to external perturbative potentials, is given by:
\begin{equation}
	\chi^{0}(\br,\br',\nu)=-i\int\frac{\text{d}{\omega}}{2\pi}e^{i\omega\delta^+}G_0(\br,\br',\omega+\nu)G_0(\br',\br,\omega).
\end{equation}
Using $\chi^0$, the dielectric function within the Random Phase Approximation (RPA) is expressed as:
\begin{equation}
	\epsilon_{\text{RPA}}(\br, \br', \omega)=\delta(\br, \br')-\int\text{d}\br''V(\br, \br'')\chi^{0}(\br'', \br',\omega).
\end{equation}
The screened Coulomb interaction within  RPA is:
\begin{equation}
	W_{\text{RPA}}(\br, \br', \omega) = \int \text{d}\br'' \left[ \epsilon^{-1}_{\text{RPA}}(\br, \br'', \omega) - \delta(\br, \br'') \right] V(\br'', \br'),
\end{equation}
where the static part has been subtracted as it is already taken into account in the HF calculations.
The correlation part of the self-energy, $\Sigma_c$, which accounts for electron correlation effects beyond HF, is then computed using:
\begin{equation}
	\Sigma_c(\br, \br', \omega) = \frac{i}{2\pi} \int d\nu \, e^{i\nu\delta^+} G_0(\br, \br', \omega + \nu) W_{\text{RPA}}(\br', \br, \nu),
\end{equation}
where $G_0$ is the HF Green's function and $W_{\text{RPA}}$ is the dynamically screened Coulomb interaction.

Then, we perform a Fourier transform from real space to reciprocal space. The matrix elements of the non-interacting charge polarizability in reciprocal space are:
\begin{equation}
	\begin{aligned}
		\chi^{0}_{\bQ\bQ'}(\btq,\nu)=&\frac{1}{N\Omega_0}\sum_{n',n,\btk}
		\left[\sum_{\lambda,\bG}
		C^*_{\lambda\bG+\bQ, n'\btk+\btq}
		C_{\lambda\bG, n\btk}\right]^*
		\left[\sum_{\lambda',\bG'}
		C^*_{\lambda'\bG'+\bQ', n'\btk+\btq}
		C_{\lambda'\bG', n\btk}\right] \nonumber\\
		&\times\left[\frac{\theta(\varepsilon_{n'\btk+\btq}-\varepsilon_F)\theta(\varepsilon_F-\varepsilon_{n\btk})}{\nu-\varepsilon_{n'\btk+\btq}+\varepsilon_{n\btk}+i\delta}
		-\frac{\theta(\varepsilon_F-\varepsilon_{n'\btk+\btq})\theta(\varepsilon_{n\btk}-\varepsilon_F)}{\nu-\varepsilon_{n'\btk+\btq}+\varepsilon_{n\btk}-i\delta}\right],\label{eq:chi0}
	\end{aligned}
\end{equation}
where \( C_{\lambda\bG, n\btk} \) are the expansion coefficients of the single-particle states in the plane-wave basis, and $\bG$, $\bG'$, $\bQ$, and $\bQ'$ denote reciprocal vectors.
The matrix form of the RPA dielectric function in reciprocal space is expressed as:
\begin{equation}
	\epsilon^{\text{RPA}}_{\bQ\bQ'}(\btq,\omega)	=\delta_{\bQ\bQ'}-V(\btq+\bQ)\chi^{0}_{\bQ\bQ'}(\btq,\omega),
\end{equation}
and the screened Coulomb interaction in reciprocal space is given by:
\begin{equation}
	W^{\text{RPA}}_{\bQ\bQ'}(\btq,\omega)=\left[ \epsilon^{-1,\text{RPA}}_{\bQ\bQ'} (\btq,\omega)-\delta_{\bQ\bQ'}\right]V(\btq+\bQ').
\end{equation}
The correlation self-energy in reciprocal space can be expressed as:
\begin{equation}
	\begin{aligned}
		\Sigma_{c}(\btk,\omega)_{nn}
		=&\frac{i}{N\Omega_0}\sum_{m,\btq}\sum_{\bG,\bG'}
		\left[ \sum_{\lambda,\bQ}C^*_{\lambda\bG+\bQ,m\btk+\btq}C_{\lambda\bQ,n\btk}\right] ^*
		\left[ \sum_{\lambda',\bQ'}C^*_{\lambda'\bG'+\bQ',m\btk+\btq}C_{\lambda'\bQ',n\btk}\right]  \nonumber\\
		&\times V(\btq+\bG)
		\int\frac{\text{d}{\nu}}{2\pi}e^{i\nu\eta}
		\frac{\left[ \epsilon^{-1,\text{RPA}}_{\bG'\bG}(\btq,\nu)-\delta_{\bG\bG'}\right] }{\omega+\nu-\varepsilon_{m\btk+\btq}+i\delta\mathrm{sgn}(\varepsilon_{m\btk+\btq}-\varepsilon_F)}.
	\end{aligned}
\end{equation}
In this formalism, the inverse dielectric function $\epsilon^{-1}_{\text{RPA}}$ plays a critical role in capturing the frequency-dependent screening of the Coulomb interaction. The QP energies are then corrected through the $GW$ self-energy, expressed as:
\begin{equation}
	\varepsilon_{n\btk}^{\text{QP}} = \varepsilon_{n\btk}^{\text{HF}} + Z_{n\btk} \, \text{Re} \, \Sigma_c(\btk, \varepsilon_{n\btk}^{\text{HF}})_{nn},
\end{equation}
where $Z_{n\btk}$ is the QP weight, accounting for interaction renormalization effects of QPs:
\begin{equation}
	Z_{n\btk} = \left[ 1 - \text{Re} \left( \frac{\partial \Sigma_c(\btk, \omega)_{nn}}{\partial \omega} \right)_{\omega = \varepsilon_{n\btk}^{\text{HF}}} \right]^{-1}.
\end{equation}
This procedure enables a more accurate determination of the single-particle energy spectrum and the QP energies by including the effects of electron correlation and dynamic screening, improving upon the HF approximation.

\begin{figure}[htb]
	\centering
	\includegraphics[width=6.4in]{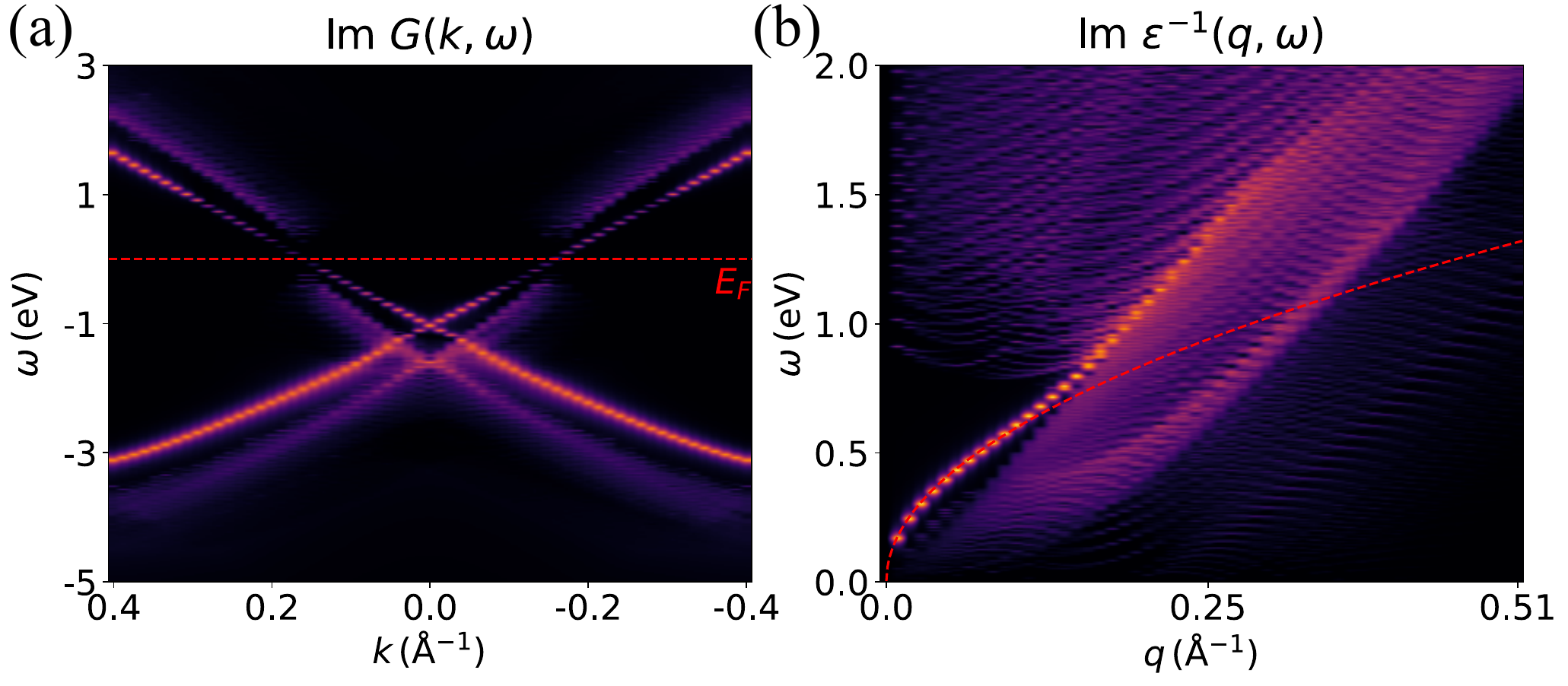}
	\caption{(a) Energy spectrum of Dirac fermions in monolayer graphene with 10\% electron doping. (b) Collective excitation spectrum.}
	\label{fig:graphene}
\end{figure}

In Supplementary Figure.~\ref{fig:graphene}(a), we present the energy spectrum of Dirac fermions in monolayer graphene after HF calculations with 10\% electron doping with respect to charge neutrality, incorporating the $GW$ self-energy correction. The main HF bands are clearly visible, along with surrounding satellite features. These satellites arise from electron-plasmon interactions, representing plasmaron or plasmon-polaron states, consistent with previous theoretical and experimental reports \cite{graphene-arpes-science10,louie-gw-graphene-prl08,zhou-plasmaron-npjqm21}. 
In Supplementary Figure.~\ref{fig:graphene}(b), we show the collective excitation spectrum corresponding to Supplementary Figure.~\ref{fig:graphene}(a). In the limit where $q$ approaches 0, it display the typical $\sqrt{q}$ dependence, indicating the presence of a gapless plasmon mode, which implies the prominent satellite feature in the $GW$ single-particle excitation spectra.

\section{Multiple plasmon pole approximation (MPA)}

The multiple plasmon pole approximation (MPA)  \cite{mpa-prb21,mpa-prb23} is an improvement over the single plasmon pole model (PPA) \cite{louie-ppa-prb86,hybertsen-ppa-prb89,needs-ppa-prl89,horsch-ppa-prb88,engel-ppa-prb93} used to approximate the dielectric function. Instead of approximating the dielectric function using a single plasmon mode, the MPA assumes the existence of multiple particle-hole collective excitations (they are all called "plasma" for simplicity). The dielectric function is then fitted using these multiple plasma (collective-excitation) modes. The approximation is given by:
\begin{equation}
	\epsilon^{-1,MPA}_{\bQ\bQ'}(\btq,\nu)-\delta_{\bQ\bQ'}
	=\sum^{N_p}_{l}\frac{2R_{l,\bQ\bQ'}(\btq)\Omega_{l,\bQ\bQ'}(\btq)}{\nu^2-\Omega^2_{l,\bQ\bQ'}(\btq)},
\end{equation}
where $N_p$ represents the number of plasmon poles, while $R$ and $\Omega$ are parameters to be determined. Specifically, $R_{l,\bQ\bQ'}(\btq)$ denotes the residue of the $l$-th plasmon mode, and $\Omega_{l,\bQ\bQ'}(\btq)$ represents its frequency. To solve for these unknown parameters, we require the values of the dielectric function at $2N_p$ different frequencies.

The advantage of the MPA is that it can simultaneously describe both plasmon modes and the continuous spectrum, avoiding the need for complicated frequency integrations, which greatly speeds up the calculation. Moreover, the accuracy of the approximation can be controlled by adjusting the number of poles in the model, allowing for a balance between computational efficiency and precision. This is especially useful for the spontaneous-symmetry breaking states such as Wigner crystal, where there are multiple collective modes such as acoustic and optical quantum phonons \cite{macdonald-wc-prb91}.

\begin{figure}[htb]
	\centering
	\includegraphics[width=6.8in]{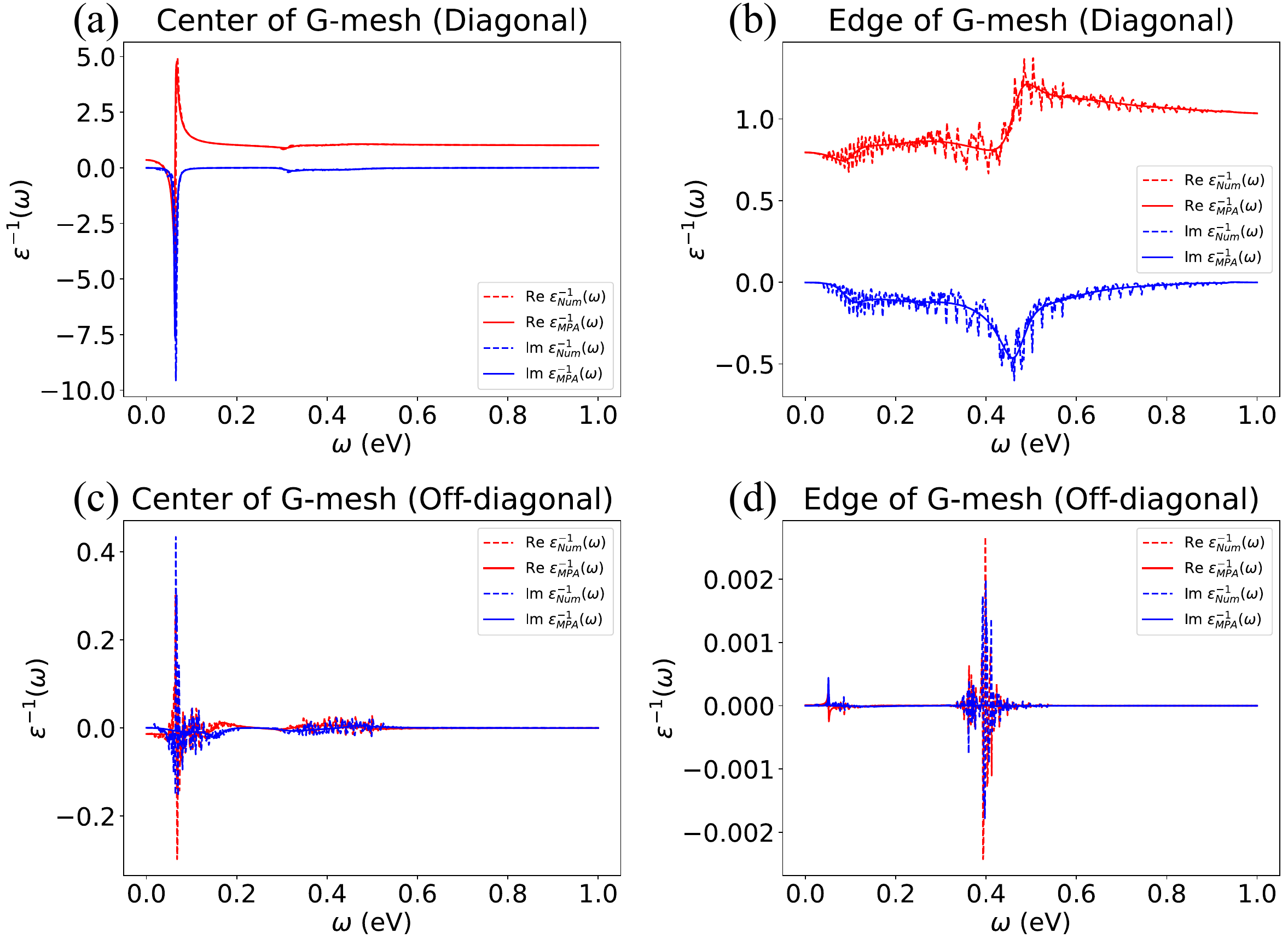}
	\caption{Comparison between the numerically calculated inverse dielectric function (dashed lines) and the MPA-fitted inverse dielectric function (solid lines). The real part is shown in red, and the imaginary part is shown in blue. (a) Small $\btq$ with $\bG = 0$ near the $\Gamma$ point, showing a prominent plasmon mode near $q = 0$. (b) Non-zero $\bG$ away from the $\Gamma$ point, entering the continuum spectrum region. (c) and (d) Off-diagonal elements near and away from the $\Gamma$ point.}
	\label{fig:MPA}
\end{figure}

Supplementary Figure~\ref{fig:MPA} shows some typical examples of the comparison between the numerically calculated inverse dielectric function (dashed lines) and the MPA-fitted inverse dielectric function (solid lines). The red and blue lines correspond to the real and imaginary parts of the inverse dielectric function, respectively. Supplementary Figure~\ref{fig:MPA}(a) presents the case of small $\btq$ with $\bG = 0$ near the $\Gamma$ point, where a clear plasmon mode is visible near $q = 0$. The MPA approximation accurately captures this plasmon mode and also describes the continuum part of the dielectric function at higher frequencies. In Supplementary Figure~\ref{fig:MPA}(b), where $\bG \neq 0$ and away from the $\Gamma$ point, the system enters the continuum spectrum region of the inverse dielectric function. The MPA approximation successfully describes the continuous spectrum using multiple plasmons, a task beyond the capability of the single plasmon pole model. Additionally, the oscillations in the numerically computed inverse dielectric function caused by $\bk$-mesh discretization (or finite-size effects) are smoothed out using MPA. Supplementary Figures~\ref{fig:MPA}(c) and \ref{fig:MPA}(d) show the off-diagonal elements of the inverse dielectric function, both near and far from the $\Gamma$ point. Although these elements are relatively small, the MPA still provides an accurate description.

Using the MPA, we can compute the correlation self-energy more efficiently without losing accuracy. The correlation self-energy $\Sigma_c$ in the MPA is given by:
\begin{align}
	\Sigma_{c}(\btk,\omega)_{nn}=&\frac{1}{N\Omega_0}\sum_{m,\btq}\sum_{\bG,\bG'}
	\left[ \sum_{\lambda,\bQ}C^*_{\lambda\bG+\bQ,m\btk+\btq}C_{\lambda\bQ,n\btk}\right] ^*
	\left[ \sum_{\lambda',\bQ'}C^*_{\lambda'\bG'+\bQ',m\btk+\btq}C_{\lambda'\bQ',n\btk}\right]  \nonumber\\
	\times &\sum^{N_p}_{l}\frac{V(\btq+\bG)R_{\bG'\bG,l}(\btq)}{\omega-\varepsilon_{m\btk+\btq}+i\delta\mathrm{sgn}(\varepsilon_{m\btk+\btq}-\varepsilon_F)+\Omega_{\bG'\bG,l}(\btq)(2f_{m\btk+\btq}-1)},
	\label{eq:self-energy}
\end{align}
where $f_{m\btk+\btq}$ represents the Fermi-Dirac distribution function.
The derivative of the correlation self-energy with respect to $\omega$ is required to determine the QP weight $Z_{n\btk}$:
\begin{align}
	\frac{\partial \Sigma_{c}(\btk,\omega)_{nn}}{\partial \omega}=&\frac{-1}{N\Omega_0}\sum_{m,\btq}\sum_{\bG,\bG'}
	\left[ \sum_{\lambda,\bQ}C^*_{\lambda\bG+\bQ,m\btk+\btq}C_{\lambda\bQ,n\btk}\right] ^*
	\left[ \sum_{\lambda',\bQ'}C^*_{\lambda'\bG'+\bQ',m\btk+\btq}C_{\lambda'\bQ',n\btk}\right]  \nonumber\\
	\times &\sum^{N_p}_{l}\frac{V(\btq+\bG)R_{\bG'\bG,l}(\btq)}{\left[ \omega-\varepsilon_{m\btk+\btq}+i\delta\mathrm{sgn}(\varepsilon_{m\btk+\btq}-\varepsilon_F)+\Omega_{\bG'\bG,l}(\btq)(2f_{m\btk+\btq}-1)\right] ^2}.
	\label{eq:self-energy-diff}
\end{align}

In the MPA, by fitting the dielectric function using multiple plasma modes, we can more accurately capture the collective excitations of the system, including both collective modes described by poles and the continuum spectra. This approach not only enhances the precision of the calculations but also significantly improves computational efficiency by reducing the complexity of frequency integration.

\section{Random phase approximation for correlation Energy}

The total energy of the system within RPA framework is given by:
\begin{equation}
	E_{\text{tot.}}=E_{\text{kin.}}+E_{\text{HF}}+E_{c}^{\text{RPA}},
\end{equation}
where $E_{\text{kin.}}$ is the kinetic energy, $E_{\text{HF}}$ is the HF energy, and $E_{c}^{\text{RPA}}$ represents the correlation energy obtained through the RPA. The inclusion of $E_{c}^{\text{RPA}}$ is crucial, as the HF approximation alone neglects correlation effects, leading to an overestimation of the tendency for symmetry-breaking states such as WC. The RPA provides a more accurate description by incorporating the effects of electron-electron interactions beyond the mean-field level.

The correlation energy in the RPA is given by \cite{fetter-book,bohm-rpa-pr53,gellmann-rpa-pr57,ren-rpa-2012}:
\begin{equation}
	\begin{aligned}
		E_c^{\text{RPA}}&=\frac{1}{4 \pi}\int_{-\infty}^{\infty} \mathrm{d} \omega \, \mathrm{Tr} \left\{ \ln \left[1-V\chi^0(i\omega)\right] + V\chi^0(i\omega) \right\} \nonumber\\
		&=\frac{1}{4 \pi}\int_{-\infty}^{\infty} \mathrm{d} \omega \sum_{\btq,\bQ,\bQ'} \left\{ \ln \left[\delta_{\bQ\bQ'} - V_{\bQ }\delta_{\bQ\bQ'}\chi^0_{\bQ'\bQ }(\btq,i\omega)\right] + V_{\bQ }\delta_{\bQ\bQ'}\chi^0_{\bQ'\bQ }(\btq,i\omega) \right\} \nonumber,
	\end{aligned}
\end{equation}
where $V$ is the bare Coulomb interaction, and $\chi^0(i\omega)$ calculated by \ref{eq:chi0} is the non-interacting charge polarizability in the imaginary frequency domain. The first line of the equation presents a general expression for the RPA correlation energy in matrix form, involving a trace over all possible interaction channels. The second line expands this expression into the momentum space, where $\btq$ and $\bQ$ denote momentum vectors, and $\chi^0_{\bQ'\bQ}(\btq,i\omega)$ are the matrix elements of the non-interacting susceptibility. 

By combining the $GW$ approximation and the RPA, we achieve a more comprehensive description of the total energy. The $GW$ approximation improves the single-particle energy spectrum, leading to more accurate values for $\chi^0(i\omega)$, while the RPA incorporates the dynamic charge fluctuation effects to provide a more reliable estimate of the correlation energy. This approach is essential for capturing the delicate balance between the FL state and other competing spontaneous symmetry-breaking phases, such as the WC state, especially in low-carrier-density interacting two-dimensional systems.

\section{Data fitting for critical Wigner-Seitz radius $r_s^*$ in 2DEG}

To determine the critical Wigner-Seitz radius $r_s^*$ for the transition between the FL state and the WC state in 2DEG system, we perform a detailed data fitting analysis. The total energies of both the WC and FL states are calculated using the $GW$+RPA framework, which incorporates both the exchange and correlation effects more accurately compared to the HF approximation.

The energy of the Wigner crystal state $E_{\text{WC}}$ is fitted using the following expression \cite{needs-wigner-qmc-prl09}:
\begin{equation}
	E_{\text{WC}} = \frac{c_1}{r_s} + \frac{c_{3/2}}{r_s^{3/2}} + \frac{c_2}{r_s^2} + \frac{c_{5/2}}{r_s^{5/2}} + \frac{c_3}{r_s^3},
\end{equation}
where the coefficients $c_1$, $c_{3/2}$, $c_2$, $c_{5/2}$, and $c_3$ are fitting parameters that capture the behavior of the WC state as a function of the Wigner-Seitz radius $r_s$. These terms account for various contributions to the energy, including kinetic, exchange, and correlation energies. The form of this equation ensures a smooth interpolation of the energy in the WC state regime.

For the Fermi liquid state, the total energy $E_{\text{FL}}$ is composed of two parts: the Hartree-Fock energy $E_{\text{FL}}^{\text{HF}}$ and the correlation energy $E_{\text{FL}}^c$ \cite{rapisarda-dqmc-1996}:
\begin{equation}
	E_{\text{FL}} = E_{\text{FL}}^{\text{HF}} + E_{\text{FL}}^c.
\end{equation}

The Hartree-Fock energy for the FL state is given by:
\begin{equation}
	E_{\text{FL}}^{\text{HF}} = \frac{1}{2r_s^2}-\frac{4\sqrt{2}}{3\pi r_s},
\end{equation}
where the first term represents the kinetic energy contribution, and the second term accounts for the exchange energy in the FL state. However, the Hartree-Fock approximation alone does not include correlation effects, which are crucial for accurately capturing the properties of the FL state.

To include the correlation effects, we use an empirical formula for the correlation energy $E_{\text{FL}}^c$:
\begin{equation}
	E_{\text{FL}}^c = a_0 \left\{1 + A x^2 \left[ B \ln \frac{x+a_1}{x} + C \ln \frac{\sqrt{x^2 + 2a_2 x + a_3}}{x}  + D\left( \arctan\frac{x+a_2}{\sqrt{a_3-a_2^2}}-\frac{\pi}{2}\right)\right]  \right\},
\end{equation}
where $x = \sqrt{r_s}$. The parameters $a_0$, $a_1$, $a_2$, and $a_3$ are fitting parameters that are determined through numerical fitting to the $GW$+RPA data. The coefficients $A$, $B$, $C$, and $D$ are given by:
\begin{equation}
	A = \frac{ 2 ( a_1 + 2a_2 )} {2a_1 a_2 - a_3 - a_1^2},
\end{equation}
\begin{equation}
	B = \frac{1}{a_1} - \frac{1}{a_1 + 2a_2},
\end{equation}
\begin{equation}
	C = \frac{a_1 - 2a_2}{a_3} + \frac{1}{a_1 + 2a_2},
\end{equation}
\begin{equation}
	D = \frac{F-a_2 C}{\sqrt{a_3 - a_2^2}},
\end{equation}
\begin{equation}
	F = 1 +  \left( 2a_2 - a_1 \right) \left( \frac{1}{a_1 + 2a_2} - \frac{2a_2}{a_3} \right).
\end{equation}

These expressions are derived to provide an accurate representation of the correlation energy in the FL state, taking into account the complex interaction effects in the 2DEG. The $GW$+RPA method gives a more precise calculation of $\chi^0$, leading to a reliable estimation of the correlation energy.

By fitting the total energies $E_{\text{WC}}$ and $E_{\text{FL}}$ obtained from the $GW$+RPA calculations, we identify the critical Wigner-Seitz radius $r_s^*$ where the energies of the WC and FL states intersect. This intersection marks the transition point between the gapless Fermi liquid state and the charge-gapped Wigner crystal state. In our calculations for the conventional 2DEG system, this critical value is found to be $r_s^* \sim 19.2$, significantly improving upon the values obtained from the HF or HF+RPA calculations. This refined critical radius demonstrates the importance of accurately including correlation effects through the $GW$+RPA framework when studying phase transitions in interacting electron systems.

\section{More results about $n$-order Dirac fermion models for $n$=2, 3, 4, 5, and 6}
\subsection{HF results}
\begin{figure}[htb]
	\centering
	\includegraphics[width=7.0in]{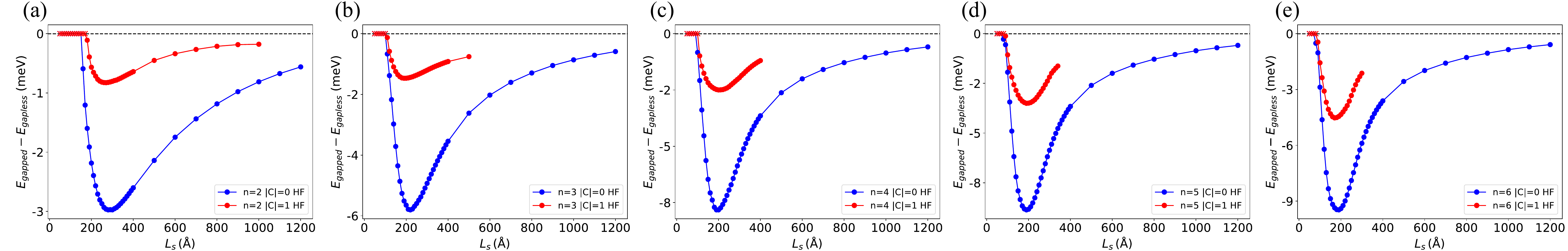}
	\caption{(a)-(e) The HF results for the WC condensation energy as a function of lattice constant $L_s$ for $n$-order Dirac fermion models with $n = 2$ to $6$, respectively.}
	\label{fig:HFphase}
\end{figure}
Supplementary Figure~\ref{fig:HFphase} shows the HF results for the WC condensation energy as a function of lattice constant $L_s$ for $n$-order Dirac fermion models with $n = 2$ to $6$. For $n=2$ to $6$ layers (corresponding to Supplementary Figures~\ref{fig:HFphase}(a)-(e)), the HF ground states for all cases are the trivial WC states. Additionally, there exists a metastable anomalous Hall crystal (AHC) state at certain values of $L_s$ for each case. However, the Chern-number-1 ground states cannot stably exist at lower electron densities, as indicated by the cutoffs at the right end of the red lines in the figures, where the Chern-number-1 metastable states under HF self-consistent calculation converge to Chern-number-0 ground states. As the number of layers increases, the condensation energies of both types of WC states at the same $L_s$ would increase in amplitude, indicating that the WC states become more stable with higher layer numbers within HF framework. This is due to the fact that systems with more layers exhibit higher density of states near Fermi energy, which further enhances interaction effects and stabilizes the WC states.

\begin{figure}[htb]
	\centering
	\includegraphics[width=7.0in]{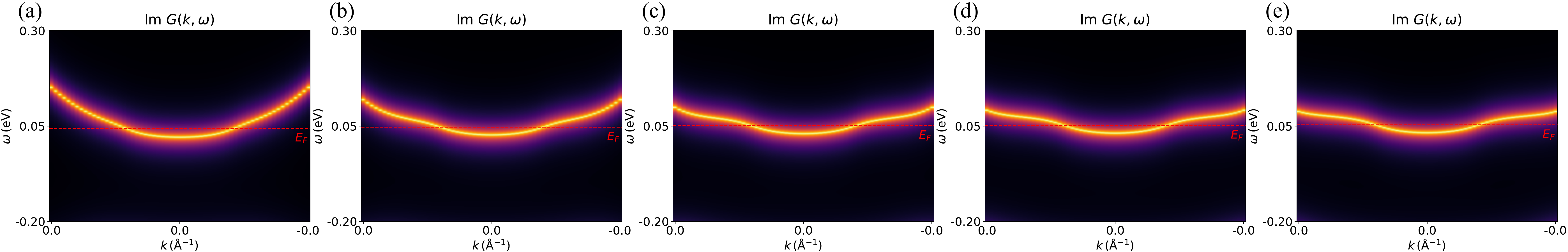}
	\caption{(a)-(e) The HF single-particle energy spectra of the FL state for $n$-order Dirac fermion models with $n = 2$ to $6$ at $L_s=200\,\text{\AA}$, respectively. The red dashed lines mark the Fermi energies.}
	\label{fig:HF_FL}
\end{figure}
Supplementary Figures~\ref{fig:HF_FL}(a)-(e) show the HF single-particle energy spectra of the FL state at for $n$-order Dirac fermion models with $n = 2$ to $6$ at $L_s=200\,\text{\AA}$, respectively. As the number of layers increases from $n=2$ to $n=6$, the low-energy bands become increasingly less dispersive, indicating stronger correlation effects with higher $n$. For lower $n$, such as in Supplementary Figure~\ref{fig:HF_FL}(a) and (b), the bands near the Fermi surface retain a relatively more dispersive character. However, as seen in Supplementary Figure~\ref{fig:HF_FL}(c)-(e), the bands flatten further, showing the enhanced influence of electron-electron interactions as the layer number increases.

\begin{figure}[htb]
	\centering
	\includegraphics[width=7.0in]{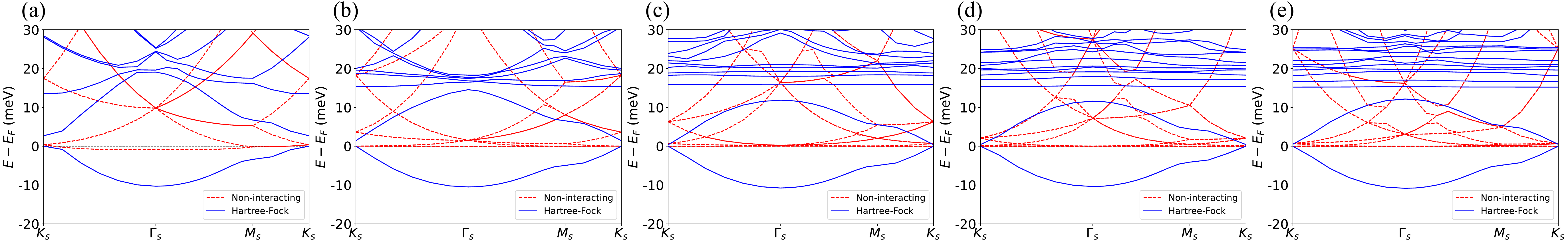}
	\caption{(a)-(e) The HF band structures for the AHC state for $n$-order Dirac fermion models with $n = 2$ to $6$ at $L_s=200\,\text{\AA}$, respectively. The gray dashed lines mark the Fermi energies.}
	\label{fig:HFband_C1}
\end{figure}
Supplementary Figures~\ref{fig:HFband_C1}(a)-(e) present the HF band structures for the AHC state for $n$-order Dirac fermion models with $n = 2$ to $6$ at $L_s = 200\,\text{\AA}$, respectively. The red dashed lines indicate the non-interacting results, where it can be observed that as $n$ increases, the low-energy bands become significantly flatter. The blue solid lines show the HF results, where an increase in $n$ leads to a gradual reduction in the bandwidth before $n=4$, and the band structures barely vary after $n=4$. 

\begin{figure}[htb]
	\centering
	\includegraphics[width=7.0in]{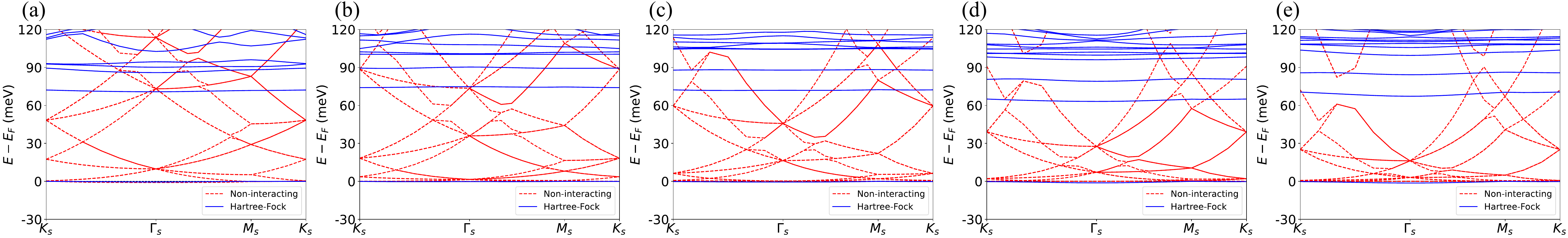}
	\caption{(a)-(e) The HF band structures for the trivial WC state for $n$-order Dirac fermion models with $n = 2$ to $6$ at $L_s=200\,\text{\AA}$, respectively.}
	\label{fig:HFband_C0}
\end{figure}
Supplementary Figures~\ref{fig:HFband_C0}(a)-(e) display the HF band structures for the trivial WC state for $n$-order Dirac fermion models with $n = 2$ to $6$ at $L_s = 200\,\text{\AA}$, respectively, in comparison with the Chern-number-1 bands shown in Supplementary Figure~\ref{fig:HFband_C1}. As $n$ increases, a key difference emerges: the low-energy bands in the Chern-number-0 state are more flat compared to those in the Chern-number-1 state, indicating a relatively stronger interaction-induced renormalization. In contrast to the Chern-number-1 state, the bandwidth of the Chern-number-0 state does not increase significantly due to interactions. This indicates a stronger tendency for electron localization in the Chern-number-0 state. This observation is consistent with the real-space charge distribution discussed in the main text, where electrons in the Chern-number-0 state form a more localized charge pattern.

\subsection{$GW$+RPA Results}

\begin{figure}[htb]
	\centering
	\includegraphics[width=7.0in]{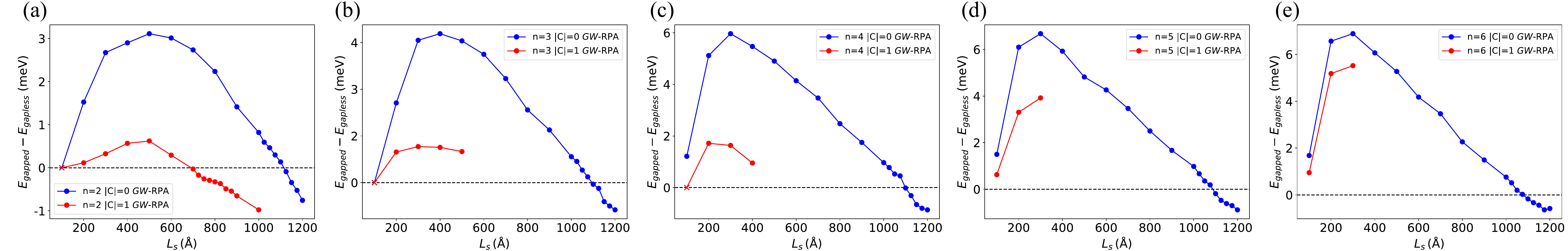}
	\caption{(a)-(e) The $GW$+RPA results for the WC condensation energy as a function of lattice constant $L_s$ for $n$-order Dirac fermion models with $n = 2$ to $6$, respectively.}
	\label{fig:GWphase}
\end{figure}
Compared to the HF ground-state results in Supplementary Figure~\ref{fig:HFphase}, the $GW$+RPA approach yields significantly different results. For the trivial WC state, similar to the 2DEG case, the more accurate $GW$+RPA method further enhances the effect of RPA correlation energy, which mitigates the overestimated exchange energy that stabilizes the WC state. This adjustment causes the phase transition for the trivial WC to occur at lower electron densities, with the layer separation increasing from approximately $L_s\sim 100\,\text{\AA}$ to $\sim1100\,\text{\AA}$ across various layers. For the AHC, the average energy gap above and below the Fermi surface is significantly smaller than in the trivial WC, leading to a substantially larger RPA correlation energy. As shown in the results from the $GW$+RPA calculations, the total energy of the AHC remains consistently lower than that of the trivial WC. Specifically, in the case of $n=2$, our calculations indicate a AHC ground state for $L_s$ values above $700\,\text{\AA}$ (corresponding to a critical density $\sim2.4\times10^{10}\text{cm}^{-2}$). These findings illustrate that the correlation energy provided by the $GW$+RPA approach is crucial in Dirac-fermion systems, effectively mitigating the exchange energy's overestimation of WC stability and yielding a more accurate depiction of the ground state.

\begin{figure}[htb]
	\centering
	\includegraphics[width=7.0in]{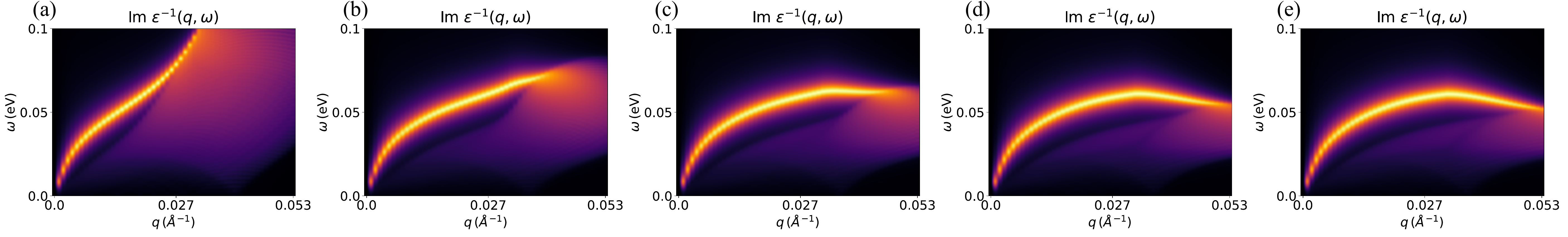}
	\caption{(a)-(e) The electron excitation spectra calculated based on the HF band for the trivial WC state for $n$-order Dirac fermion models with $n = 2$ to $6$ at $L_s=200\,\text{\AA}$, respectively.}
	\label{fig:GWcoll}
\end{figure}
Supplementary Figures~\ref{fig:GWcoll}(a)-(e) show the collective excitation spectra for $n$-order Dirac fermion models with $n = 2$ to $6$ at $L_s = 200\,\text{\AA}$, respectively, calculated based on the HF energy bands. 
Here we focus on the most significant intraband low-energy collective excitations. 
Besides the particle-hole continuum, in the limit where $q$ approaches 0, all layers display the typical $\sqrt{q}$ dependence, indicating the presence of a gapless plasmon mode. However, the bandwidth of these modes differs notably in the $n=2, 3, 4$ systems. As the layer number increases to $n=5$ and $n=6$, the HF bands become extremely flat, and the collective excitation spectra exhibit a weakly dispersive
branch over a certain range of wavevector. This behavior implies the presence of a prominent satellite feature in the $GW$ single-particle excitation spectra.

\begin{figure}[htb]
	\centering
	\includegraphics[width=7.0in]{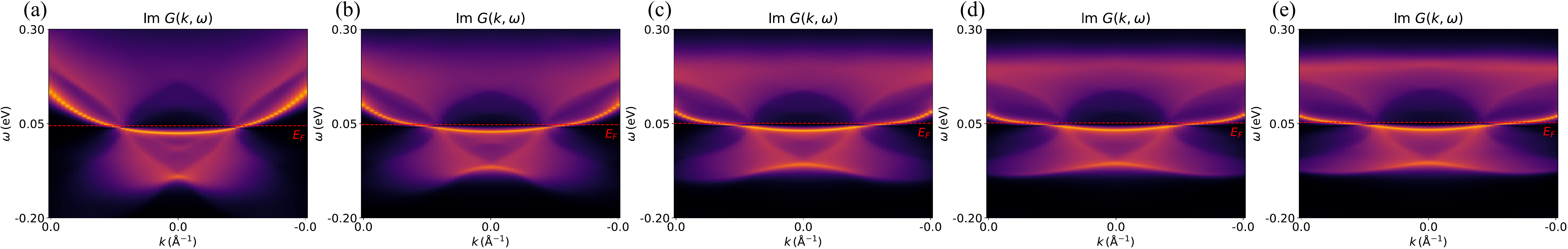}
	\caption{(a)-(e) The $GW$-calculated single-particle energy spectra for the FL state for $n$-order Dirac fermion models with $n = 2$ to $6$ at $L_s=200\,\text{\AA}$, respectively.}
	\label{fig:GW_FL}
\end{figure}
Supplementary Figures~\ref{fig:GW_FL}(a)-(e) show the $GW$-calculated single-particle energy spectra for the FL state for $n$-order Dirac fermion models with $n = 2$ to $6$ at $L_s = 200\,\text{\AA}$, respectively. Near the Fermi surface, distinct plasmon satellites are visible, indicating the influence of electron-plasmon interactions. Furthermore, as shown in Supplementary Figure~\ref{fig:GWcoll}, with increasing layer number, a progressively flatter and nearly momentum-independent spectral feature emerges below the Fermi surface. This flat feature arises from the interaction between electrons and plasmons, leading to the formation of plasmarons or plasma-polarons. Similar features in the single-particle spectra due to electron-plasmon couplings have also been reported in carrier-doped monolayer graphene \cite{graphene-arpes-science10,louie-gw-graphene-prl08,zhou-plasmaron-npjqm21}.  The appearance of this feature becomes more pronounced with smaller bandwidth in higher-layer systems, resulting from the flatter bands in the corresponding non-interacting models.

\begin{figure}[htb]
	\centering
	\includegraphics[width=7.0in]{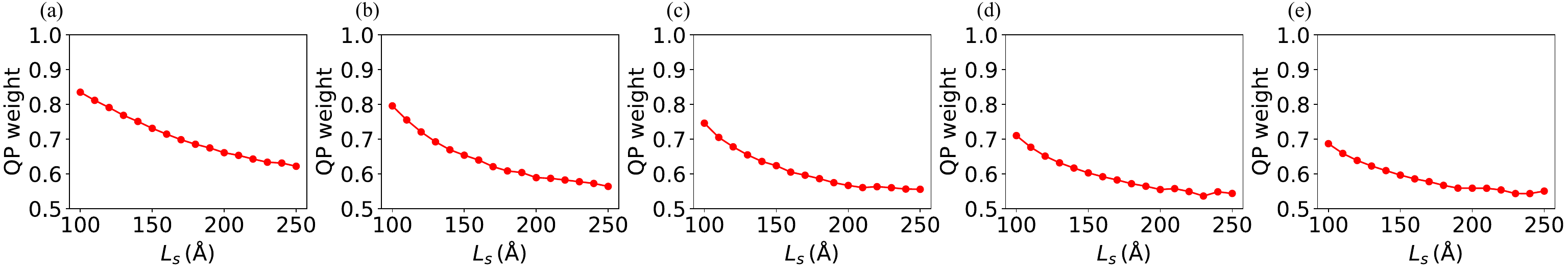}
	\caption{(a)-(e) The $GW$-calculated QP weight for the FL state for $n$-order Dirac fermion models with $n = 2$ to $6$, respectively.}
	\label{fig:QP}
\end{figure}
Supplementary Figures~\ref{fig:QP}(a)-(e) show the $GW$-calculated QP weight for the FL state for $n$-order Dirac fermion models with $n = 2$ to $6$, respectively. As the layer number $n$ increases, the QP weight at the same $L_s$ gradually decreases, indicating a reduction in the coherent part of the spectral weight. This reduction reflects the increasing interaction effects in higher-layer systems, consistent with the enhanced electron-plasmon interactions observed in the corresponding $GW$ single-particle spectra.

\begin{figure}[htb]
	\centering
	\includegraphics[width=7.0in]{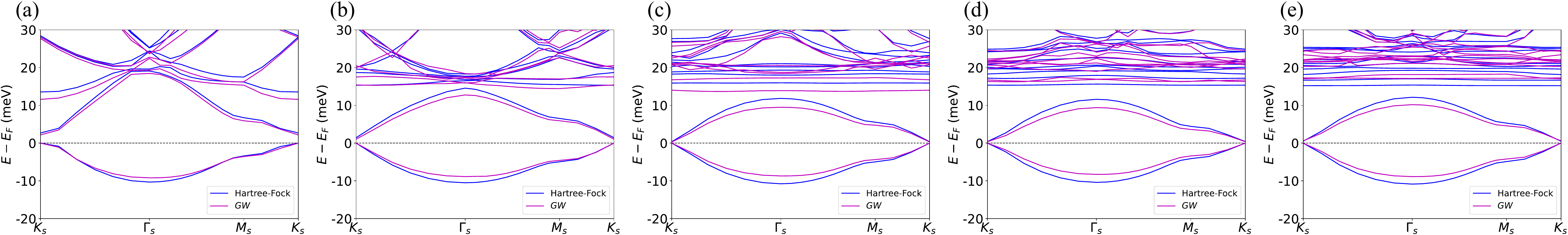}
	\caption{(a)-(e) The $GW$-calculated single-particle energy spectra for the AHC state for $n$-order Dirac fermion models with $n = 2$ to $6$ at $L_s=200\,\text{\AA}$, respectively.}
	\label{fig:GWband_C1}
\end{figure}
\begin{figure}[htb]
	\centering
	\includegraphics[width=7.0in]{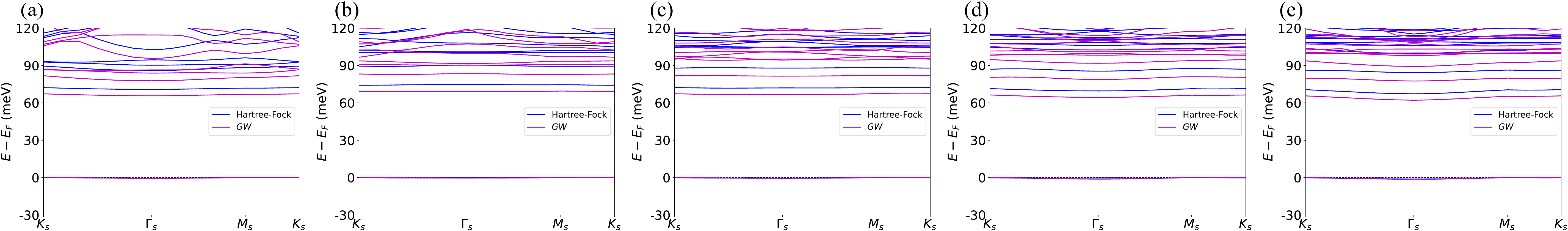}
	\caption{(a)-(e) The $GW$-calculated single-particle energy spectra for the trivial WC state for $n$-order Dirac fermion models with $n = 2$ to $6$ at $L_s=200\,\text{\AA}$, respectively.}
	\label{fig:GWband_C0}
\end{figure}
Supplementary Figures~\ref{fig:GWband_C1}(a)-(e) and Supplementary Figures~\ref{fig:GWband_C0}(a)-(e) show the $GW$ quasi-particle band structures for the AHC and trivial WC states for $n$-order Dirac fermion models with $n = 2$ to $6$ at $L_s = 200\,\text{\AA}$, respectively, compared to the HF results in Supplementary Figures~\ref{fig:HFband_C1}(a)-(e) and Supplementary Figures~\ref{fig:HFband_C0}(a)-(e). The $GW$ correction introduces dynamical screening of the Coulomb interactions, resulting in a more accurate calculation of the bandwidth and band gap. This correction mitigates the overestimation by the HF approximation, leading to a reduced bandwidth and a more realistic depiction of the gap.
\subsection{Trigonal warping effects in bilayer graphene (BLG)}
We continue to present results about the more realistic bilayer graphene two-band model by considering the trigonal warping effect. Supplementary Figures~\ref{fig:tri_phase}(a) displays the HF ground state phase diagram with trigonal warping included. Compared to the Dirac fermion model without trigonal warping, the condensation energies of both types of WC ground states are significantly enhanced by the exchange effect. Additionally, the energy of the trivial WC consistently remains lower than that of the AHC ground state. Supplementary Figures~\ref{fig:tri_phase}(b) illustrates the ground state phase diagram obtained from GW+RPA calculations. It is evident that when $L_s\lessapprox 750\,\AA$, the trivial WC remains the system's ground state. However, when $L_s\lessapprox 750\,\AA$, the AHC state becomes energetically stabilized over the trivial WC, which is attributed to the lower correlation energy achieved by AHC through dynamical charge fluctuations.
\begin{figure}[htb]
	\centering
	\includegraphics[width=3.0in]{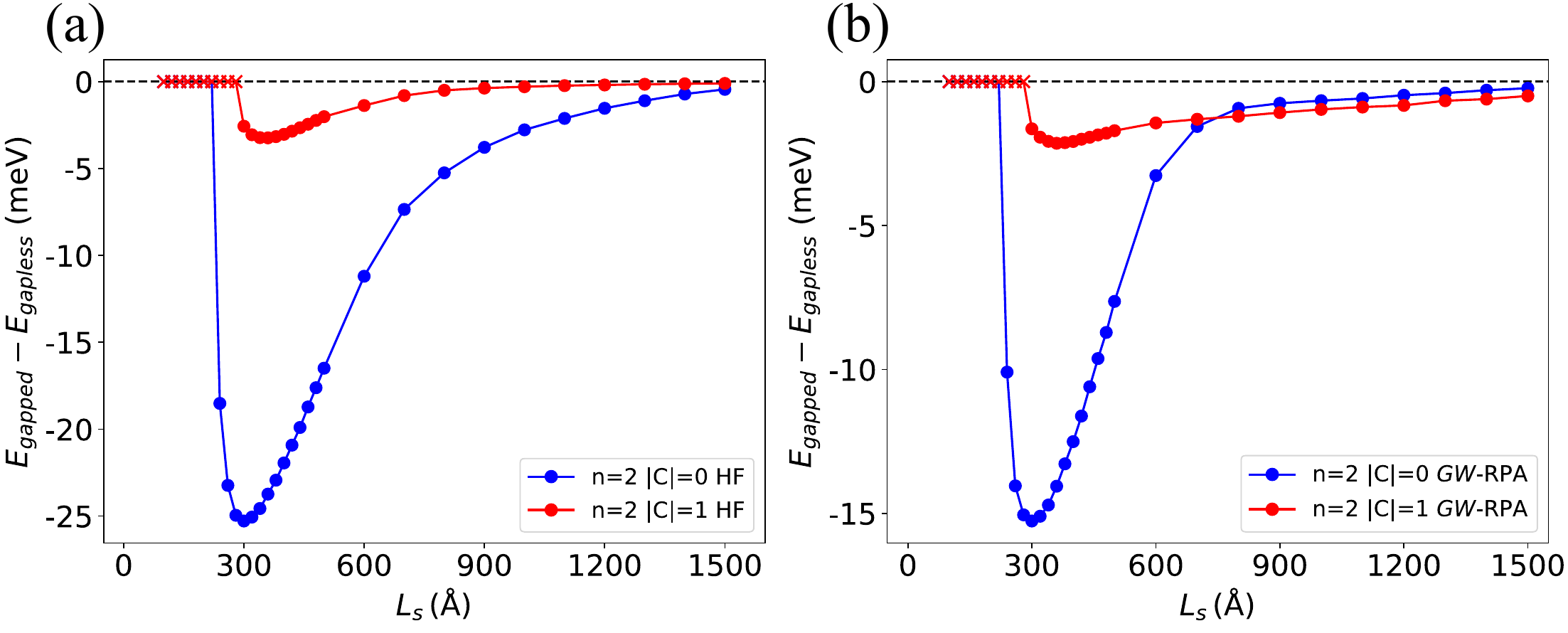}
	\caption{(a)-(b) The HF and $GW$+RPA results for the WC condensation energy as a function of lattice constant $L_s$ for bilayer graphene two band model, respectively.}
	\label{fig:tri_phase}
\end{figure}
\subsection{Summary}
In this section, we have presented a comprehensive analysis of $n$-order Dirac fermion models for $n = 2$ to $6$ and the BLG two-band model with trigonal warping, exploring the ground state under both HF and the more advanced $GW$+RPA calculations. The HF results indicate that, with an increasing layer number, the WC state becomes more stable due to enhanced electron-electron interactions, with the trivial WC as the ground state in this context. However, the inclusion of RPA correlation energy in the $GW$+RPA framework reveals that correlation effects shift the WC phase transition point to a lower electron density and lower the total energy of the AHC below that of the trivial WC. This effect is particularly evident in the $n = 2$ system, where, at low electron densities, the correlation energy enables the AHC to surpass the trivial WC as the ground state. In the BLG two-band model with trigonal warping, the AHC is favored over the trivial WC at low density due to lower correlation energy gained from dynamical charge fluctuations. This finding underscores the indispensable role of correlation effects described by the $GW$+RPA approach in the study of the WC phase transition.
	
\end{widetext}
\end{document}